\documentclass{aastex62}
\usepackage{natbib}

\usepackage{subfigure}

\usepackage{amsmath}
\usepackage[outdir=./]{epstopdf}
\usepackage{multirow}
\usepackage{booktabs}
\usepackage{graphicx}
\usepackage{hyperref}
\usepackage{bm}

\usepackage{url}

\begin{document}
\title{Determine the Masses and  Ages  of  Red Giant Branch Stars  from Low-resolution LAMOST Spectra Using DenseNet}

\author{   Xuejie Li }
   \affil{School of Space Science and Physics, Shandong University, Weihai, 264209,  Shandong, China}

 \author{Yude Bu $^\star$}
  
  \affil{ School of Mathematics and Statistics, Shandong University, Weihai, 264209,   Shandong, China}
   \email{  $^\star$ buyude@sdu.edu.cn}
\author{  Jianhang Xie}
   \affil{School of Mechanical, Electrical \& Information Engineering, Shandong University, Weihai, 264209,  Shandong, China}

    \author{  Junchao Liang}
  \affil{ School of Mathematics and Statistics, Shandong University, Weihai, 264209,   Shandong, China}

    \author{  Jingyu Xu}
  \affil{ School of Mathematics and Statistics, Shandong University, Weihai, 264209,   Shandong, China}

 \begin{abstract}
  We propose a new model to determine the ages and masses of red giant branch (RGB) stars from the low-resolution large sky area multi-object fiber spectroscopic telescope (LAMOST) spectra. \textbf{The ages of RGB stars are difficult to determine using classical isochrone fitting techniques in the Hertzsprung--Russell diagram}, because isochrones of  RGB stars are tightly  crowned. With the help of the asteroseismic method, we can determine \textbf{the masses and ages of RGB stars} accurately. Using the ages derived from the asteroseismic method, we train a deep learning model based on DenseNet to calculate the ages of RGB stars directly from their spectra. We  then apply this model to determine the ages of {512\,272} RGB stars from \textbf{LAMOST DR7} spectra (see \url{http://dr7.lamost.org/}). The results show that our model can estimate the ages of RGB stars from low-resolution spectra with an accuracy of 24.3$\%$. The results on the open clusters M 67, Berkeley 32, and NGC 2420  show that our model performs well in estimating the ages of RGB stars. Through comparison, we find that our method performs better than other methods in determining the ages of RGB stars. The proposed method  can be used in the stellar parameter pipeline of upcoming large surveys such as 4MOST, WEAVES, and MOONS.

\end{abstract}

\section{Introduction}
	Red giant branch (RGB) stars usually have high luminosity and can be detected from a long distance, which are good probes for studying the structure of the Milky Way. Hence, the accurate estimation of the parameters of RGB stars plays a significant role in understanding the synthesis of star clusters and the evolutionary history of the Milky Way. \textbf{The age of an RGB star reflecting its evolutionary stage is one of the key parameters, but it is considerably difficult to estimate directly.}  \\ \indent
	 The most widely used method to measure ages of large samples of stars includes determining their location in the Hertzsprung--Russell diagram (HRD) and  comparing them with theoretical isochrones simultaneously \citep{edvardsson1993chemical, feltzing2001solar, da2006basic, bergemann2014gaia, sanders2018isochrone}. However, this method works well only for main-sequence turn-off stars and those on the subgiant branch. With respect to the RGB stars, this method is not appropriate, \textbf{because stellar isochrones of different ages have similar positions in the HRD, and small uncertainties in temperature can lead to large uncertainties in age}. \textbf{\citet{soderblom2010ages} summarizes the methods of estimating the ages of stars and discusses robustness, accuracy, and some model-dependent or empirical methods for stellar age estimation.}\\ \indent
	Because the main-sequence lifetime of an RGB star is \textbf{closely} related to its mass, once the  mass is accurately calculated, the age can be  estimated from stellar evolutionary tracks \citep{pinsonneault2018second}. For an RGB star, once we have its asteroseismic parameters, we can accurately determine its mass. There are two global asteroseismic parameters: $\Delta\nu$, the frequency spacing between two modes of the same spherical degree and consecutive radial order, and $\nu_{\mathrm{max}}$, \textbf{the frequency at maximum acoustic power} \citep{kjeldsen1994amplitudes}, which can be applied to infer mass with the oscillation data from seismic scaling relations. A series of space missions such as Convection, Rotation, and planetary Transits
satellite  \citep[CoRoT;][]{baglin2008corot, sharma2016stellar}, Kepler \citep{borucki2010kepler,2008IAUS..249...17B}, Transiting Exoplanet Survey Satellite \citep[TESS;][]{ricker2014transiting}, and K2 \citep{howell2014k2} have yielded measurements of seismic parameters for thousands of RGB stars \citep{hekker2009characteristics, hekker2011characterization, stello2013asteroseismic, stello2015oscillating, stello2017k2}.  \textbf{However, it is difficult to obtain $\Delta\nu$, one of the critical parameters to estimate masses, of a large sample of stars. Hence more space-based missions are required for large-scale asteroseismology. Thus, it is necessary to develop alternative methods to determine  the ages of many RGB stars.}  \\ \indent
	\citet{ness2016spectroscopic} proposed a data-driven approach using   $The\ Cannon$ \citep{ness2015cannon} to determine stellar masses and ages of  RGB stars from APOGEE DR12 with accuracies of $\sim$ 0.07 and $\sim$ 0.2 dex (40\%), respectively. \citet{ho2017masses} used the method proposed  in \citet{martig2016red} to infer stellar masses and ages for 230\,000 RGB stars from large sky area multi-object fiber spectroscopic telescope (LAMOST) catalog with accuracies of 0.08 and 0.2 dex, respectively. \citet{wu2018mass} estimated masses and ages of 6\,940 RGB stars with asteroseismic parameters deduced from Kepler photometry and stellar atmospheric parameters derived from LAMOST spectra. The typical uncertainty of mass is a few percent, and that of age is $\sim$ 20$\%$. Recently, \citet{wu2019ages} proposed a model based on the KPCA method to estimate the ages of RGB stars. They first trained the model by taking the LAMOST-Kepler giant stars with asteroseismic parameters and the LAMOST-TGAS subgiant stars based on isochrones as training sets, and then 
applied the model to the 640\,986 RGB stars from the LAMOST Galactic Spectroscopic Survey (DR4) to obtain their ages and masses. Experiments show that the age and mass of RGB sample stars with signal-to-noise ratio (SNR) $>$ 30 given by the model have a median error of 30\% and 10\%, respectively. \\ \indent	
	Currently, machine learning methods, especially deep learning models, are popular in many fields. 
	\citet{hon2018deep} used a one-dimensional (1D) convolutional neural network to classify the evolutionary state of oscillating red giants efficiently into red giant branch stars and helium-core burning stars by recognizing visual features in their asteroseismic frequency spectra. 
	\citet{zhang2019938} used StarNet (a deep neural network) to estimate stellar parameters ($T_\text{eff}$, log g, [M/H]), $\alpha$-elements, as well as C and N abundances from LAMOST spectra, using stars in common with the APOGEE survey as training dataset, yielding uncertainties of 45 K for $T_\text{eff}$, 0.1 dex for $\log g$, 0.05 dex for [M/H], 0.03 dex for $[\alpha/M]$, 0.06 dex for [C/M], and 0.07 dex for [N/M]. \citet{2019MNRAS.483.3255L} applied an artificial neural network (ANN)  to infer 22 stellar elemental
abundance labels for both high- and
low-SNR spectra simultaneously from APOGEE DR14.  \\ \indent
	In this work, we use a deep learning method, densely connected convolutional network (DenseNet), to estimate masses and ages of the common  RGB stars from the LAMOST and Kepler surveys. DenseNet is a new type of deep learning architecture based on a convolutional neural network, whose main feature is that each layer connects every other layer in a feed-forward fashion \citep{huang2017densely}. The results of DenseNet show that it can estimate the  masses of RGB stars with a median error of {6.5$\%$} and their ages with an accuracy of 24.3$\%$. This study provides a new method for estimating the ages of RGB stars, which can be used to estimate the ages of RGB stars obtained by upcoming large surveys. \\ \indent
    The remainder of this paper is organized as follows. 
    Section \ref{Data} introduces the data used in the experiment. Section \ref{model} introduces basic concepts and theories of our de-noising model and the estimation model based on a modified version of DenseNet, followed by Section \ref{estimation} describing  the overall  algorithm containing data preprocessing, de-noising, and DenseNet-BC training. In Section \ref{results}, the final experimental results are presented with the accuracy and precision analysis. Section \ref{discussion} introduces the classification model for RGB selection, the test results on open clusters, and two comparisons, including the accuracy of different machine learning models and the age estimation derived from different methods.
    
\section{Data}\label{Data}
\begin{figure}[htbp]
    \centering
    \subfigure[]{
        \includegraphics[width=0.45\textwidth]{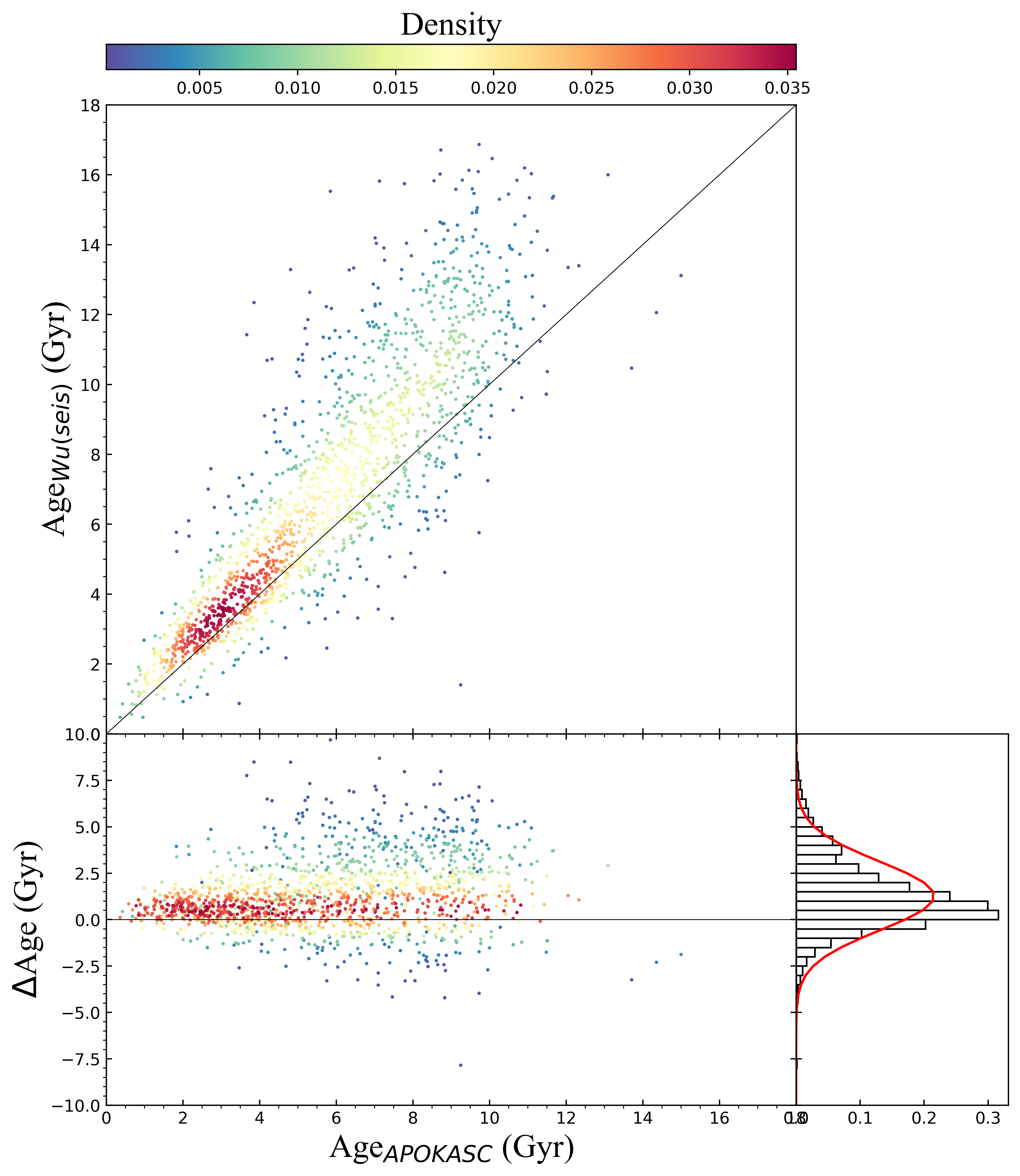}
        \label{fig:age_apo_wu(seis)}
    }
    \subfigure[]{
	\includegraphics[width=0.45\textwidth]{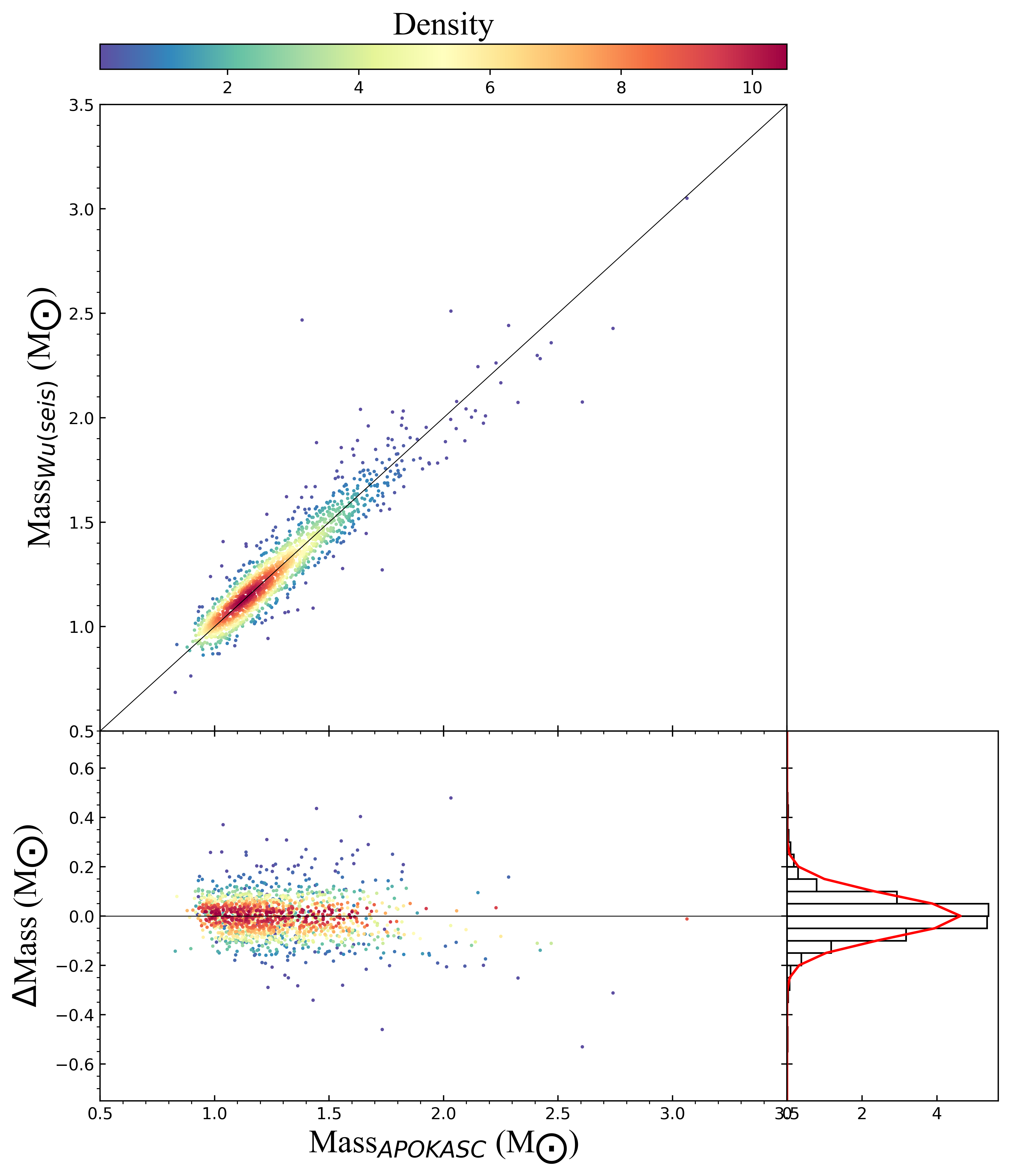}
        \label{fig:mass_apo_wu(seis)}
    }
    \caption{Comparison between the asteroseismic age/mass given by \citet{wu2018mass} and the age/mass provided by the APOKASC-2 catalog \citep{pinsonneault2018second}. The left panel shows the residual distribution in the age comparison, and the right panel shows that for the mass comparison.}
\end{figure}

 The data used by us for age estimation \textbf{are from \citet{wu2018mass}}, and they consist of the 3\,352 common RGB  stars of LAMOST survey and Kepler survey. They are selected from the LAMOST-Kepler  project, which  was initiated to use the large sky area multi-object fiber spectroscopic telescope (LAMOST) to make spectroscopic follow-up observations of the targets in the field of the Kepler mission \citep{2015ApJS..220...19D,2018ApJS..238...30Z}. LAMOST, located at the Xinglong Observatory in the northeast of Beijing, China \citep{2012RAA....12.1197C}, is a special reflecting Schmidt telescope that  has a wide field ($5^\circ$ in diameter) and a large effective aperture (4--6 m, depending on the pointing altitude
and hour angle).
 LAMOST can collect 4\,000 spectra simultaneously in a single exposure, and is a powerful spectroscopic survey telescope for wide-field and
large-sample astronomy. After its pilot survey from October 2011 to June 2012 and a five-year regular survey from September 2012 to June 2017, LAMOST has collected low-resolution (R $\sim$1800) optical
spectra ($\lambda$ 3700--9000 $\mathring{A}$ ) of more than 9\,000\,000 stars down to a limiting magnitude of r $\sim$18 mag in a contiguous sky area ($-10^\circ$ to +90$^\circ$ declination).    
\\ \indent
\textbf{
To determine the masses and ages of these stars, \citet{wu2018mass} used the following steps:
\begin{enumerate}
    \item[(1)] Filter out red clump (RC) stars that have an uncertain amount of mass loss and exert a negative impact on the age estimation. The selected RGB stars consist of two parts: (i) All the stars  with the asteroseismic parameters $\nu_{\text{max}} > 120\ \mu\text{Hz and}\ \nu_{\text{max}} < 12 \ \mu\text{Hz}$. (ii) 
    Some stars with asteroseismic parameters in the range of $12 \ \mu\text{Hz} <\nu_{\text{max}} < 120\ \mu\text{Hz}$ are selected referring to other literature (\citealp{bedding2011gravity,stello2013asteroseismic,mosser2014mixed,vrard2016period,elsworth2017new}), because RGB and RC stars coexist in this interval.
    \item[(2)]  Determine the masses with the corrected scaling relation as in Equation \eqref{scale1}--\eqref{scale2} \citep{sharma2016stellar}. 
\begin{equation}
    \frac{M}{M_\sun}=\left(\frac{\Delta \nu}{f_{\Delta\nu}\Delta \nu_\sun}\right)^{-4}\left(\frac{\nu_{\mathrm{max}}}{\nu_{\mathrm{max},\sun}}\right)^3 \left(\frac{T_{\text{eff}}}{T_{\text{eff},\sun}}\right)^{1.5}
    \label{scale1}
\end{equation}
\begin{equation}
    \frac{R}{R_\sun}= \left(\frac{\Delta \nu}{f_{\Delta\nu}\Delta \nu_\sun}\right)^{-2}\left(\frac{\nu_{\mathrm{max}}}{\nu_{\mathrm{max},\sun}}\right) \left(\frac{T_\text{eff}}{T_{\text{eff},\sun}}\right)^{0.5}
    \label{scale2}
\end{equation}
Here, $T_{\text{eff},\sun}=$ 5\,777 K, $\nu_{\mathrm{max},\sun}=$ 3\,090 $\mu$Hz, and $\Delta \nu_\sun$ =135.1 $\mu$Hz. $f_{\Delta\nu}$ is a modification factor generated from the ASFGRID code \citep{sharma2016stellar}. $T_{\text{eff}}$ is given by the LAMOST Stellar Parameter Pipeline at Peking University \citep[LSP3;][]{xiang2015lamost}. 
\item[(3)] Determine the ages from Yonsei--Yale ($Y^2$) isochrones with the Overshoot scheme and $\alpha$-enhancement utilizing the Bayesian fitting method similar to \citet{xiang2017ages}.
The initial mass function (IMF) used in the fitting method from \citet{kroupa2001variation} is shown in Equation \eqref{eq:IMF},
\begin{equation}
    \label{eq:IMF}
    \xi(m) \propto m^{-\alpha},
\end{equation}
where  $\alpha=0.3$ for $m < 0.08\ M_{\sun}$, $\alpha=1.3$ for $0.08 < m < 0.5\ M_{\sun}$, and $\alpha=2.3$ for $m > 0.5\ M_{\sun}$ \citep{xiang2017ages}. 
The input parameters are masses, $\log g$, $T_{\text{eff}}$, [Fe/H], and [$\alpha$/Fe], which are derived  with the LSP3 method,  and the typical precisions are approximately 0.1 dex for $\log$ g, 100 K for $T_{\text{eff}}$, 0.1 dex for [Fe/H], and better than 0.05 dex for [$\alpha$/Fe].
\end{enumerate}
}
\textbf{
The determined age is robust owing to precise mass estimation, and the typical relative age errors are 15--35 $\%$, with a median value of 25$\%$. 
\citet{wu2018mass} compared the results of different isochrones, including the Dartmouth Stellar Evolution Program \citep[DSEP;][]{dotter2008dartmouth} and the PAdova and TRieste Stellar Evolution Code \citep[PARSEC;][]{bressan2012parsec}. Ages estimated with the DSEP isochrones are comparable to those estimated with the $Y^2$ isochrones, with a mean difference of 6\% and a dispersion of 8\%. Ages estimated with the PARSEC isochrones are also consistent with those estimated with the $Y^2$ isochrones.
Note that none of the three isochrones considered the effects from stellar rotation, metal diffusion, and magnetic fields, which may have had some impact on the age determination.
Figure \ref{fig:age_apo_wu(seis)} and Figure \ref{fig:mass_apo_wu(seis)} show our comparison of the masses and ages derived from the above process with those from APOKASC-2 \citep{pinsonneault2018second}. The mass estimation compared to APOKASC-2 shows good consistency. 
The age from \citet{wu2018mass} has a mean difference of 25\% from those   from \citet{pinsonneault2018second}, which  uses interpolation in the BeSPP grid at the input mass, surface gravity, [Fe/H], and [$\alpha$/Fe]. }

\begin{figure*}[!h]
	\centering
	\includegraphics[width=1\linewidth]{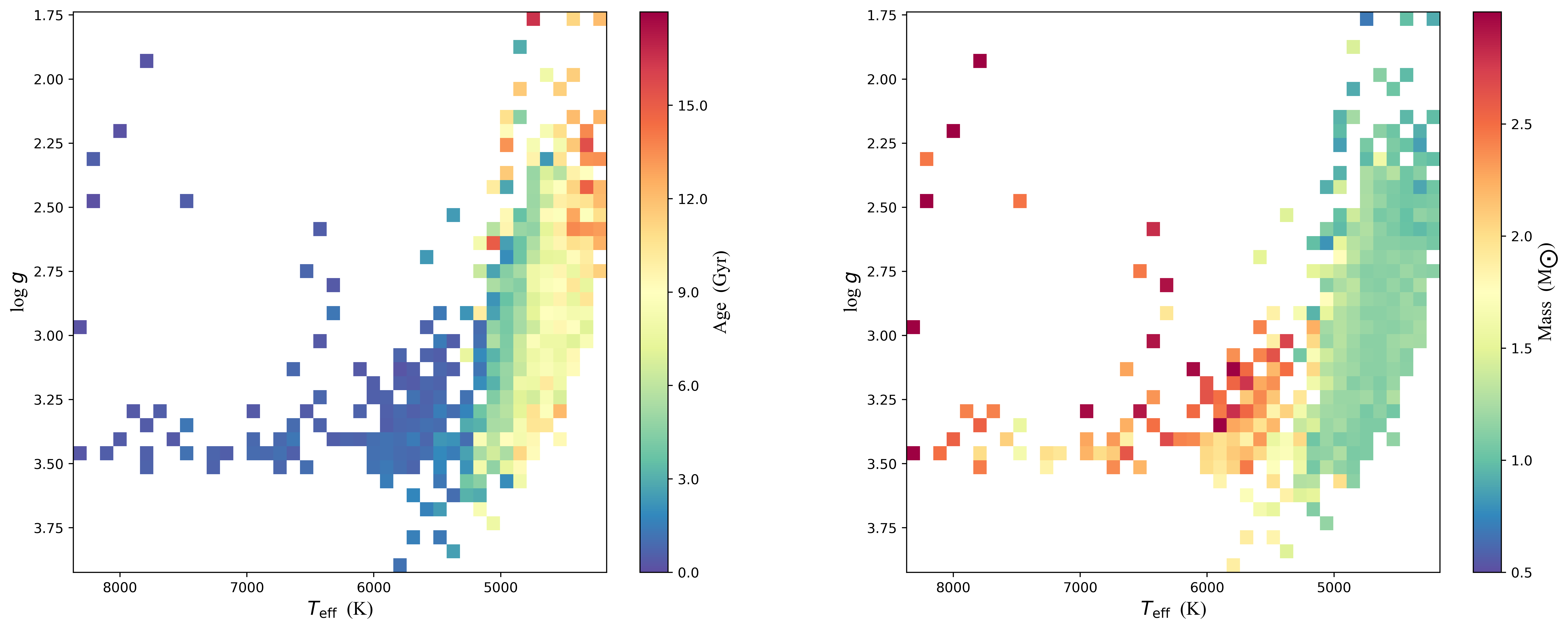}
	\caption{Distributions of median stellar ages (left-hand panel) and masses (right-hand panel) in the $T_{\mathrm{eff}}$-$\log\ g$ plane. The color bar represents the median age and mass of each bin.}
	\label{fig:stellar_paras}
\end{figure*}

\textbf{We then select stars with age errors smaller than 40\% or mass errors smaller than 15\%. After cross-match with LAMOST DR6, our dataset contains 3\,352 stars, with a median mass error of 7\% and a median age error of 20\%.
To reduce the systematic errors caused by the boundary effects of the selection method \citep{xiang2017estimating}, we add another 188 LAMOST-TGAS subgiant stars and main-sequence turn-off stars to the dataset.The masses and ages
 of these stars are determined by the same method introduced above. In total, our dataset contains 3\,540 stars with precise seismic mass and age estimates. Figure \ref{fig:stellar_paras} shows the distribution of these RGB stars in the HRD. The surface gravities range from 1.75 to 3.92 dex and the effective temperatures vary from 4\,140 to 8\,370 K.}

\section{Methodology}
\textbf{In the first part of this section, we briefly describe the concepts and basic theories of a denoising model based on the wavelet transform and a regression model based on DenseNet modified for spectral data. In the second part, we introduce the method of data processing and describe how to apply them to the models in the estimation of ages and masses.}
\subsection{Model}\label{model}
	\subsubsection{Denoising Model Based on Wavelet Transform}\label{sec:denoising model}
		 Each stellar spectrum consists of continuous spectra, spectral lines, and noise. The noise in the spectrum may lead to large errors in age estimation. Hence, to improve the accuracy of age estimation, we should denoise the spectra using an appropriate denoising method. The wavelet transform \citep{meyer1992wavelets} is a signal analysis method that has the ability to represent the local properties of a signal in both the time and frequency domains, and has been widely used in signal denoising. To increase the SNR and improve the prediction accuracy, we denoise the spectra by applying the wavelet transform. 
		
		First, we briefly introduce the basic theory of the wavelet transform. Assume we have a function $\psi(t)$ with finite energy, whose Fourier transform is $\Psi(\omega)$. When it satisfies the \textbf{admissibility} condition 
		\begin{equation*}
			c_{\psi} = 2\pi\int_{-\infty}^{\infty}\frac{|\Psi(\omega)|^2}{|\omega|}\mathrm{d}\omega < \infty, 
		\end{equation*}
	  $\psi(t)$ is called the mother wavelet. Define
	  	\begin{equation*}
			\psi_{a,b}(t) = \frac{1}{\sqrt{a}}\psi(\frac{t-b}{a}) \qquad  
		\end{equation*}
	 where the parameter $a\in R$ represents the scale and the parameter $b\in R$ represents the time shift. Suppose that $s(t)$ is a spectrum, then the continuous wavelet transform (CWT) of $s(t)$ is defined as
		\begin{equation*}
			\omega(a,b)=\int_{-\infty}^{\infty}\psi^*_{a,b}(t)s(t)\mathrm{d}t
		\end{equation*}
		 where  $\psi^*_{a,b}(t)$ is the complex conjugate of $\psi_{a,b}(t).$  The inverse wavelet transform is defined by
		\begin{equation*}
			s(t) = \frac{1}{c_{\psi}}\int_{-\infty}^{\infty}\int_{-\infty}^{\infty} \frac{1}{a^2} \omega(a,b)\psi_{a,b}(t)\mathrm{d}a\mathrm{d}b.
		\end{equation*}
		When we discretize the continuous scale parameter $a$ and the time shift parameter $b$ by 
		\begin{align*}
			a = a_0^m,\ b = nb_0 \qquad m,n \in \textbf{Z},  
			\end{align*}
		then $ \psi_{m,n}(t) = \frac{1}{\sqrt{a_0^m}}\psi(\frac{t-nb_0}{a_0^m}) $. 
	 The  discrete wavelet transform (DWT) and its inverse transform are defined by
		\begin{align*}
			\omega_{m,n} = \int_{-\infty}^{\infty}\psi^*_{m,n}(t)s(t)\mathrm{d}t \\
			s(t) = k_{\psi}\sum_m\sum_n \omega_{m,n}\psi_{m,n}(t)
		\end{align*}
		where $k_{\psi}$ is the normalization coefficient.

\begin{figure*}[!h]
			\centering
			\includegraphics[width=1\linewidth]{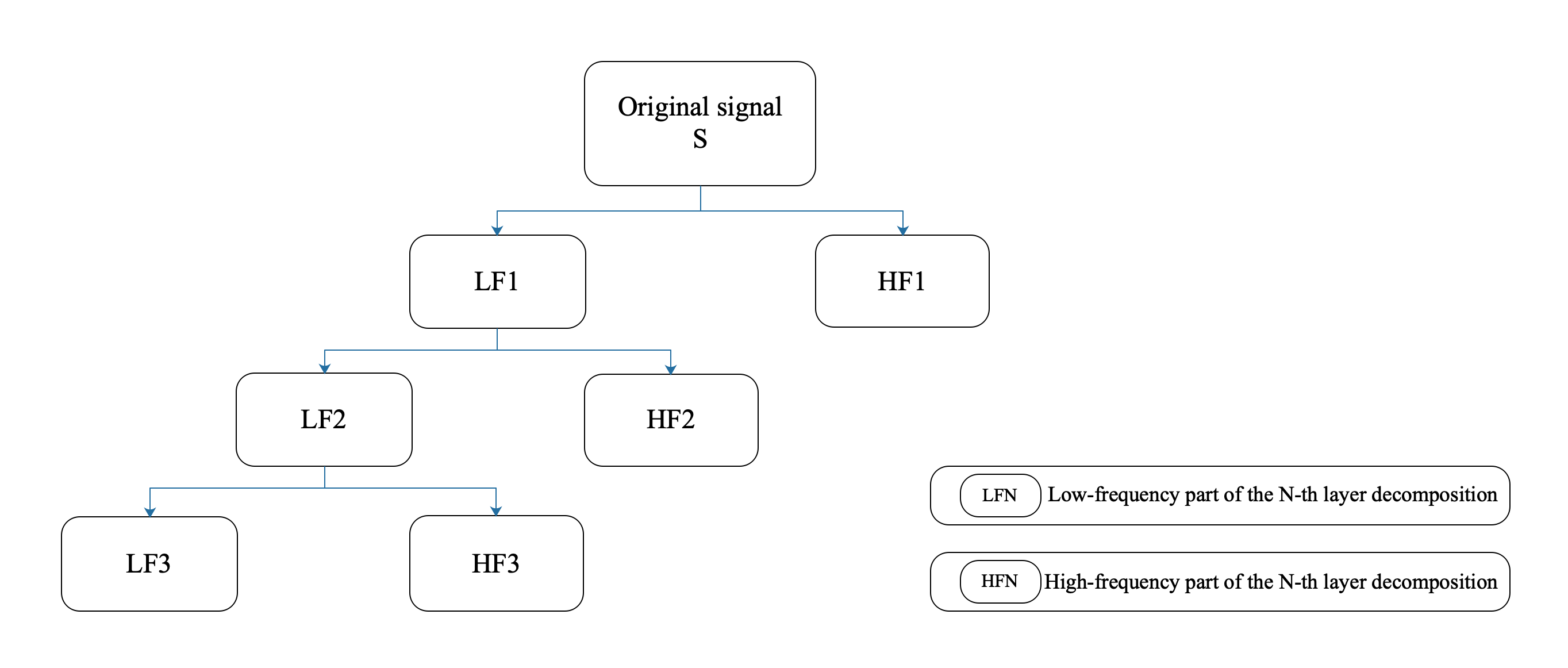}
			\caption{Structure of the three-layer decomposition example.}
			\label{fig:MRA}
		\end{figure*}

	  \citet{mallat1989theory} introduced the concept of multiresolution analysis (MRA) and  proposed a method to construct an orthogonal wavelet and a fast algorithm of orthogonal wavelet analysis. According to wavelet theory, a low-resolution signal can be linearly expressed by a high-resolution signal, such that we can see that the space tensed by the high-resolution signal must contain the space tensed by the low-resolution. Suppose that space $L_M~(M=1,2,\cdots,N,\cdots)$ is the subspace of $L^2(R)$ and satisfies
		$$L^2(R)=L_0{\supset} L_1\supset L_2\cdots L_N\cdots. $$
		For each $M$, we have $L_M=L_{M+1}{\oplus} H_{M+1}$. 
		Then, we have 
		$$L^2(R)=L_0=L_1\oplus H_1=L_2\oplus H_2\oplus H_1=L_3\oplus H_3\oplus H_2\oplus H_1 =\cdots $$
		 Next, we define projection operator $HF_M: L^2(R)\rightarrow H_M$ and $LF_M: L^2(R)\rightarrow L_M$.
		Then, for each spectrum $S\in L^2(R)$, we have
		$$S=LF_1(S)+HF_1(S)=LF_2(S)+HF_2(S)+HF_1(S)=LF_3(S)+HF_3(S)+HF_2(S)+HF_1(S),$$
		We  call  $LF_i(S)$  the low-frequency of the $i$-th decomposition, and $HF_i(S)$ the high-frequency of the $i$-th decomposition.
		 A three-layer decomposition example of MRA is illustrated   in Figure \ref{fig:MRA}.

	 \citet{donoho1995noising} introduced soft thresholding and hard thresholding \textbf{to estimate  the wavelet coefficients, $\omega(a,b)$.} The threshold is the estimated noise level. Values larger than the threshold are regarded as signal, and smaller values are noise. 
		
		Based on the basic theory of DWT, Mallat's algorithm and soft thresholding, we use the following three steps to  denoise the spectrum:
		\begin{itemize}
			\item Step 1: According to Mallat's algorithm, decompose 1D spectrum data and determine the number of composition layers $N$ by Equation \eqref{eq1}, in which $N_d$ is the length of the input data and $l_f$ is the length of the wavelet decomposition filter. \textbf{Then, we obtain the wavelet coefficients $\omega_{LF,i}^k$ (the $i$-th wavelet coefficients in $LFk$) and $\omega_{HF,j}^k$ (the $j$-th wavelet coefficients in $HFk$).}
				\begin{equation}\label{eq1}
					N = \log_2\left(\frac{N_d}{l_f-1}\right)
				\end{equation}
			\item Step 2: Quantize the high-frequency wavelet coefficients $\omega_{HF,j}^k$ through the soft threshold function, which is defined by Equation \eqref{eq2}, where $T$ is the threshold, to get quantized high-frequency wavelet coefficients $\hat\omega_{HF,j}^k$. This step is critical to filter the noise component.
				\begin{align}\label{eq2}
					\hat\omega_{HF,j}^k=
					\begin{cases}
						0 &\qquad |\omega_{HF,j}^k| < T, \\			 	\text{sign}(\omega_{HF,j}^k)\cdot(|\omega_{HF,j}^k| - T) &\qquad |\omega_{HF,j}^k| \geq T.
					\end{cases}
				\end{align}
			\item Step 3: Reconstruct the signal \textbf{with $\omega_{LF,i}^N$ and $\hat\omega_{HF,j}^k,\ k=1,2,\cdots, N$}, and  obtain  access to the denoised spectral data.
		\end{itemize}

	\subsubsection{Regression Model Based on DenseNet-BC}
\textbf{Deep neural networks, which are collections of artificial neurons that are connected to one another in the form of real number weights, are typically used to simulate specific mappings. A fully connected network consists of multiple layers, where neurons are connected to the previous layer. The input of each layer is calculated by a combination of weights and values from the previous layer, and the output comes from a nonlinear function (activation function) of the input, which promises complex approximate mappings. A convolutional neural network, a variation of the deep neural network, is apt at extracting features from the input data, where the weights in described above now act as convolutional filters. The basic idea of a convolutional neural network is the multilayer mechanism in which alternating convolution and pooling layers propagate information.}
\\ \indent 
DenseNet is a high-performance image classification convolutional neural network  in the domain of computer version introduced by \citet{huang2017densely}. Compared with other deep learning methods, DenseNet has the following  advantages: (i) it is capable of alleviating the vanishing-gradient and exploding-gradient problems that occur in deep convolutional neural networks;
	 (ii) it strengthens the connection between the former layers and the latter layers, which successfully stimulate feature reuse;
	 (iii) \textbf{it considerably reduces the number of parameters to be tuned in the network.}
			\begin{figure*}[!h]
				\centering
				\includegraphics[width=1\linewidth]{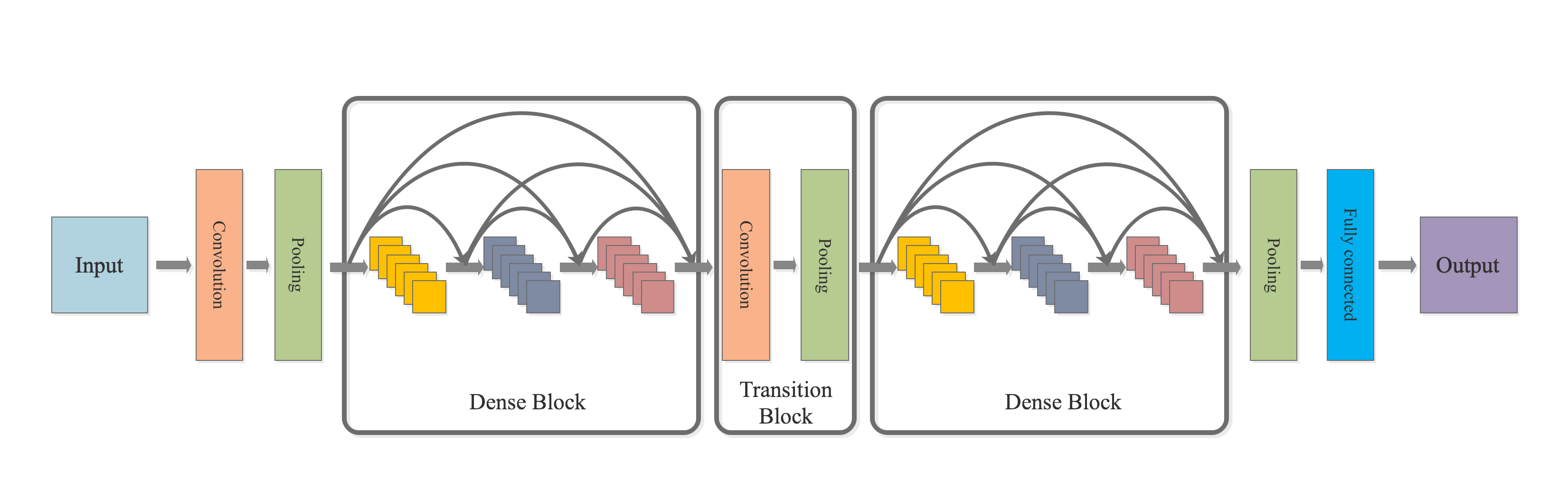}
				\caption{Fundamental structure of DenseNet. The grey arrows indicate the flow of feature maps in the network.}
				\label{fig:fdDN}
			\end{figure*}

\textbf{
The basic DenseNet is composed of dense blocks and transition blocks (see Figure \ref{fig:fdDN}) that contain different layers, including
convolutional (Conv), max pooling (Pooling), fully connected (FC), dropout, and batch normalization (BN) layers. 
The convolutional layer can effectively extract features by sliding the filter across the input. The output of each layer consists of a set of subimages called feature maps,
whose size is the same as that of the input.
The pooling layer can downsample the output of the convolutional layer and is usually used to reduce the complexity of the network and improve the spatial invariance with respect to the position of features. The pooling layer does not alter the number of feature maps, but it can change their size. 
The fully connected layer holds composite and aggregated information from all the convolutional layers, representing the feature vector for the input, which is then further used for classification or regression. 
The dropout layer is used to alleviate overfitting and improve generalization capacity. 
Batch normalization \citep{ioffe2015batch} is used to pull the skewed distribution back to the standardized distribution, such that the input value of the activation function falls in the area where the activation function is more sensitive to the input.
All the convolutional layers are followed by the rectified linear unit activation function \citep[ReLU;][]{glorot2011deep}. 
}

		A dense block contains a set of convolutional layers, each of which is connected to every layer before it. Hence, if a dense block has $L$ layers, there will be $\frac{L(L+1)}{2}$ connections. Moreover, every Conv in every dense block has the same number of channels, called the growth rate $k$. A transition block is a module that connects two dense blocks and decreases the size of feature maps. It is made of a BN, a $(1,1)$ Conv, and a pooling layer. Owing to the addition of the $1\times 1$ Conv, a transition block can compress the model as well. \textbf{If the dense block prior to this transition block} outputs $m$ feature maps, it could produce $\theta m$ feature maps, where $\theta \in (0,1]$ is the compression rate. 

			\begin{figure*}
						\includegraphics[width=1\linewidth]{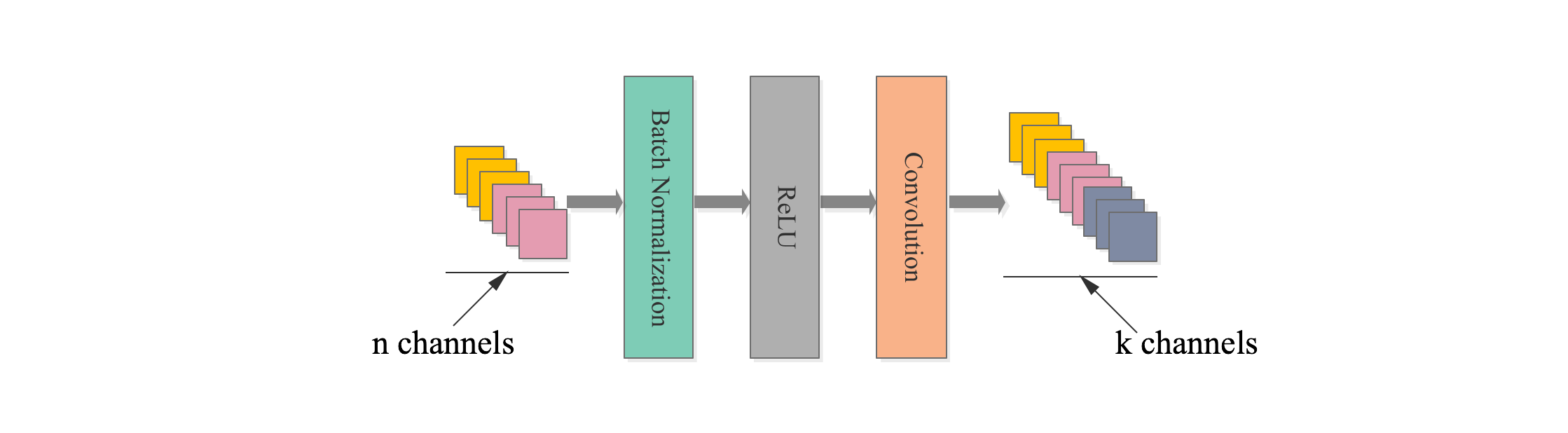}
						\caption{Structure of a basic dense block.}
						\label{fig:DN_DB}
			\end{figure*}
				\begin{figure*}
						\includegraphics[width=1\linewidth]{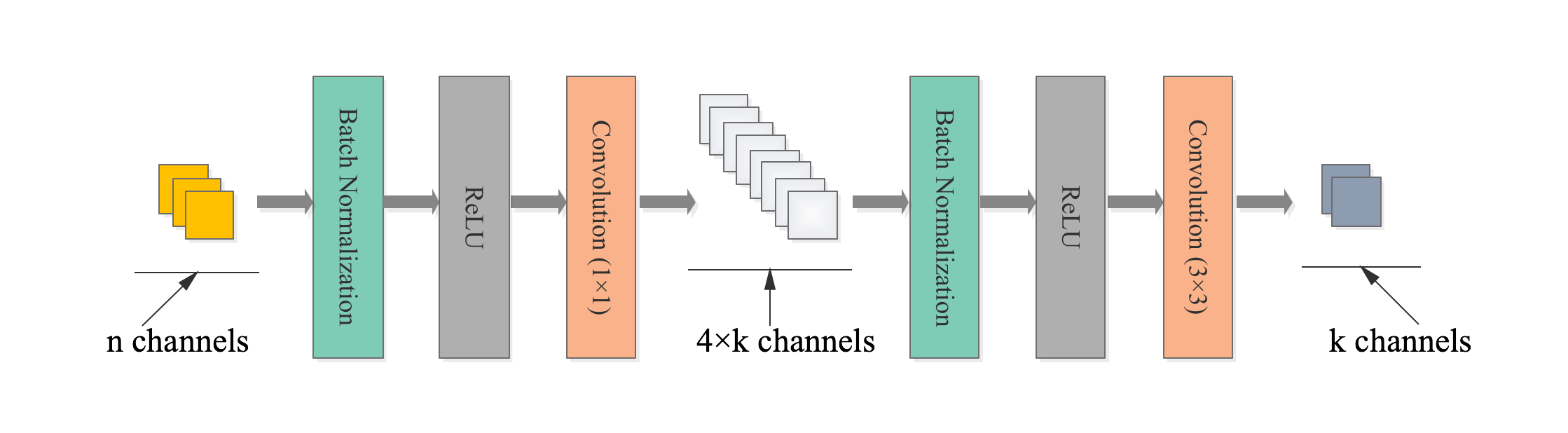}
						\caption{Structure of dense block with bottleneck layers. $1\times 1$ Conv gives $4\times k$ feature maps.}
						\label{fig:DN_B_DB}
			\end{figure*}
\begin{figure*}
				\centering
				\includegraphics[width=1\linewidth]{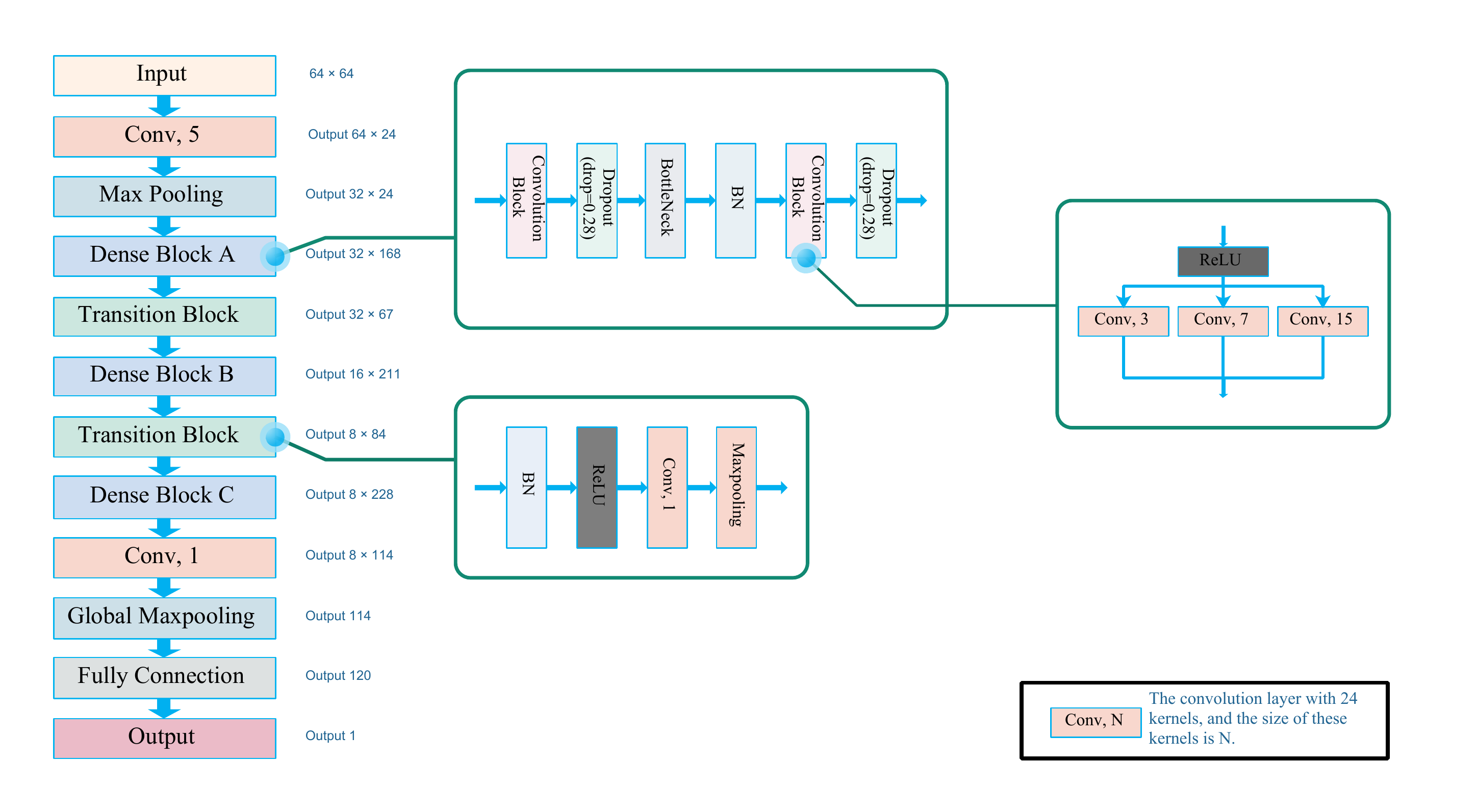}
				\caption{Architecture of the constructed 1D DenseNet-BC and the details of dense, transition, and convolution blocks. The size and the number of feature maps output by each module are demonstrated in the form of $Output\ size\times channels$. The introduction of a convolution layer is in the bottom right corner of the figure.}
				\label{fig:my DN}
			\end{figure*}

			DenseNet-BC  is a variation of DenseNet. Compared with DenseNet, DenseNet-BC has some modifications:
			(i) Bottleneck layers (mainly $1\times 1$ Conv) are added into dense blocks of DenseNet to improve computational efficiency (Figure \ref{fig:DN_DB}, \ref{fig:DN_B_DB});
			(ii) a transition layer following $m$ layers of dense block generates $\theta m$ ($0<\theta<1$) output feature maps.   \\
			\indent In this study, we built a 1D DenseNet-BC model to satisfy the demand of age and mass evaluation, with some modifications. As demonstrated in Figure \ref{fig:my DN}, our model uses three dense blocks and two transition blocks. Unlike the original DenseNet-BC architecture, considering the distribution of features in spectral data are not even, we made a smaller block named convolution block with parallel construction. Each convolution block uses a different kernel size $(3, 7, 15)$ to get more features, and then concatenates the convolution results. In the transition block, the compression rate was set to  0.4.
			Additionally, a 1D Conv with a kernel size of 1 called the integration layer was arranged behind the last dense block to integrate the information, which allows the model to minimize the complexity of the fully connected layer.

\subsection{Age and Mass Estimation}\label{estimation}
	This section describes the algorithm of estimating ages and masses of RGB stars from spectral data. Because the spectra are high-dimensional data, they are difficult to deal with and most machine learning methods we used (e.g., XGBoost and artificial network) performed undesirable. 
	 To solve these problems, we first used the wavelet transform to denoise and transform the spectrum to a multichannel one. Then, we applied a deep learning framework to this complicated regression problem. 
	The data preprocessing procedure is shown below:
		
	\begin{itemize}
		\item Step 1: \textbf{Interpolate the spectral data. To ensure that the coverage of each sample is in the same range $[4000, 8096]\ \mathring{\mathrm{A}}$, we use linear interpolation to fill missing values of the spectra in this interval. Therefore, each sample in the dataset has 4096 dimensions.}
		\item Step 2: Remove false data. As shown in Figure \ref{fig:missing flux window}, there exists a window of missing flux, which can exert a negative influence on the model training. The selection rules are as follows:
			\begin{align*}
				COUNT(flux=0) &< 100
			\end{align*}
			\begin{figure}
				\centering
				\includegraphics[height=80mm,width=0.8\linewidth]{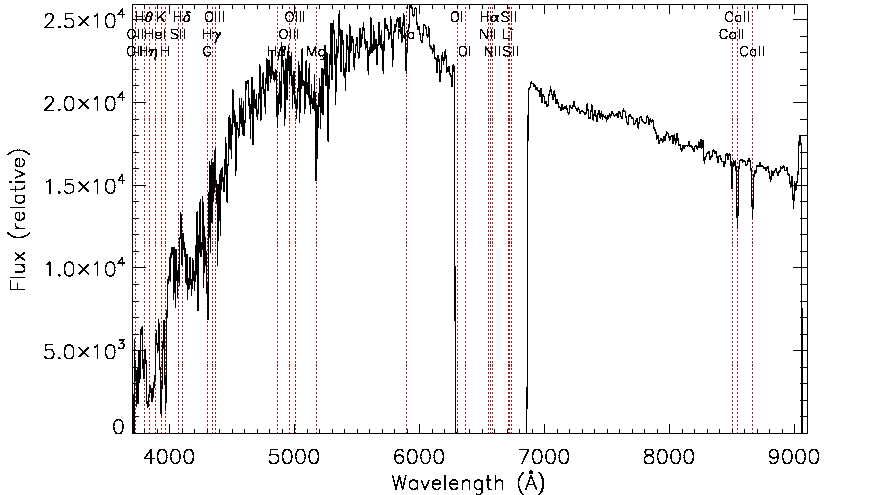}
				\caption{Spectrum with missing flux in the range of 6300--6850 $\mathring{\mathrm{A}}$ from our original dataset. This figure comes from \url{http://dr6.lamost.org/}.}
				\label{fig:missing flux window}
			\end{figure}
			
		\item Step 3: Normalize the spectral data. To improve the robustness of model, make the model focus on the spectrum construction instead of the flux, and accelerate the following process; we normalize the spectral data using Equation \eqref{eq5}.
			\begin{equation}\label{eq5}
				F = \frac{F-F_{\mathrm{min}}}{F_{\mathrm{max}}-F_{\mathrm{min}}}
			\end{equation}
		Here, $F$ is the raw spectral flux data, and $F_{\mathrm{min}}$ and $F_{\mathrm{max}}$ are the minimum and maximum values of raw spectral flux data, respectively.
		\item Step 4: Re-drop the unadaptable spectral data. Some of the spectral data that have a discernable difference from most RGB spectra cannot be put into the model because their noise component overshadows the signal component. After the calculation, the re-drop rule is as follows:
		\begin{equation*}
			std(F) < 0.005
		\end{equation*}
		\item Step 5: Denoise the spectra. We used the denoising model introduced in Section \ref{sec:denoising model}, and chose the \textbf{Haar wavelet \citep{chui2016introduction}} as the basic function, which is defined by Equation \eqref{haar} and $0.05\times \max{\{\omega_{HF,j}^k\}}$ as the threshold $T$. Figure \ref{fig:raw_vs_de-noised} shows the comparison of the raw and denoised spectra.
		\begin{figure}
			\centering
			\includegraphics[width=0.8\linewidth]{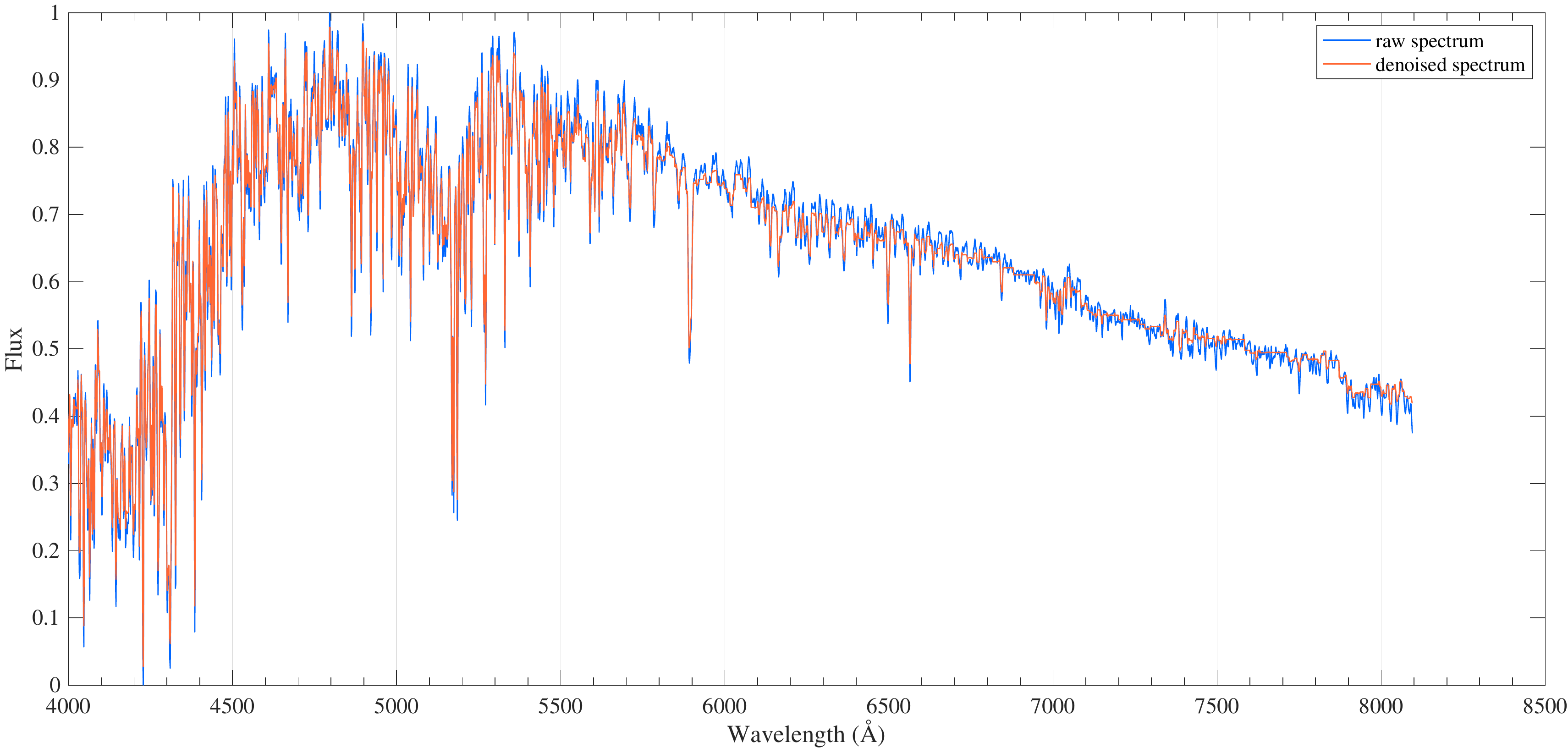}
			\caption{Comparison of the raw spectrum (in blue) and the corresponding denoised spectrum (in orange) using wavelet transform.}
			\label{fig:raw_vs_de-noised}
		\end{figure}
		\begin{equation}
		    \label{haar}
		    \psi(t)=
		    \begin{cases}
		        1\qquad & 0\leq t<\frac{1}{2}, \\
		        -1\qquad &\frac{1}{2}\leq t <1, \\
		        0\qquad &\text{otherwise}.
		    \end{cases}
		\end{equation}
		\item Step 6: 
		\textbf{Reshape the spectral data. We divide each spectrum into 64 bands on average, which is similar to the shape of multiband image data $(\text{height} \times \text{width} \times \text{channels})$. Then, we use them as the input of the 1D multichannel DenseNet-BC model. Compared with directly inputting a spectrum of length, this effectively reduces the number of parameters and saves computational costs.}
	\end{itemize}
 When training our DenseNet-BC model in age estimation, we use \textbf{Huber loss \citep{huber1992robust} in Equation \eqref{eq:huber}} as the loss function, which is a balanced way between the mean absolute error and the mean squared error. The Huber loss is defined by 
 \begin{equation}
     \label{eq:huber}
     L_{\delta}(y,\hat{y})=
     \begin{cases}
         \frac{1}{2}(y - \hat{y})^2, \qquad &\text{for}\  |y-\hat{y}|\leq\delta\\
         \delta\cdot(|y-\hat{y}|-\frac{1}{2}\delta),\qquad&\text{otherwise,}
     \end{cases}
 \end{equation}
 where $\delta$ is the parameter of Huber loss, and $y$ and $\hat{y}$ are the true and predicted values, respectively. 
Adam \citep{kingma2014adam} is used as the optimizer to update weights of network. The method of initializing weights is He uniform variance scaling initializer \citep{he2015delving}. The learning rate is 0.00015. 

\section{Results}\label{results}
	\subsection{Measure of Results}
		In this section, we report the results of DenseNet-BC model. \textbf{The preprocessed data are divided into two independent groups in a ratio of 8:2 (training set and validation set)}. 
		We use the following two metrics to evaluate the performance of the experiment:
		\begin{align*}
		&M_1=\frac{\sum\limits_{i=1}^{n}|\hat{y}_i-y_i|}{n},\\
		&M_2=\frac{\sum\limits_{i=1}^{n}|\hat{y}_i-y_i|/y_i}{n}\times100\%,
		\end{align*}
		where $n$ represents the number of samples in a dataset, and $\hat{y}_i$ and $y_i$ represent the prediction value and the true value of the $i$-th sample, respectively. $M_1$ measures mean absolute error and $M_2$ measures the mean absolute percentage error. Both measure the distance between the prediction value and the real value, such that the lower their values, the better the performance. 
		\begin{figure}[htbp]
			\centering
			\subfigure[]{
				\includegraphics[width=0.45\textwidth]{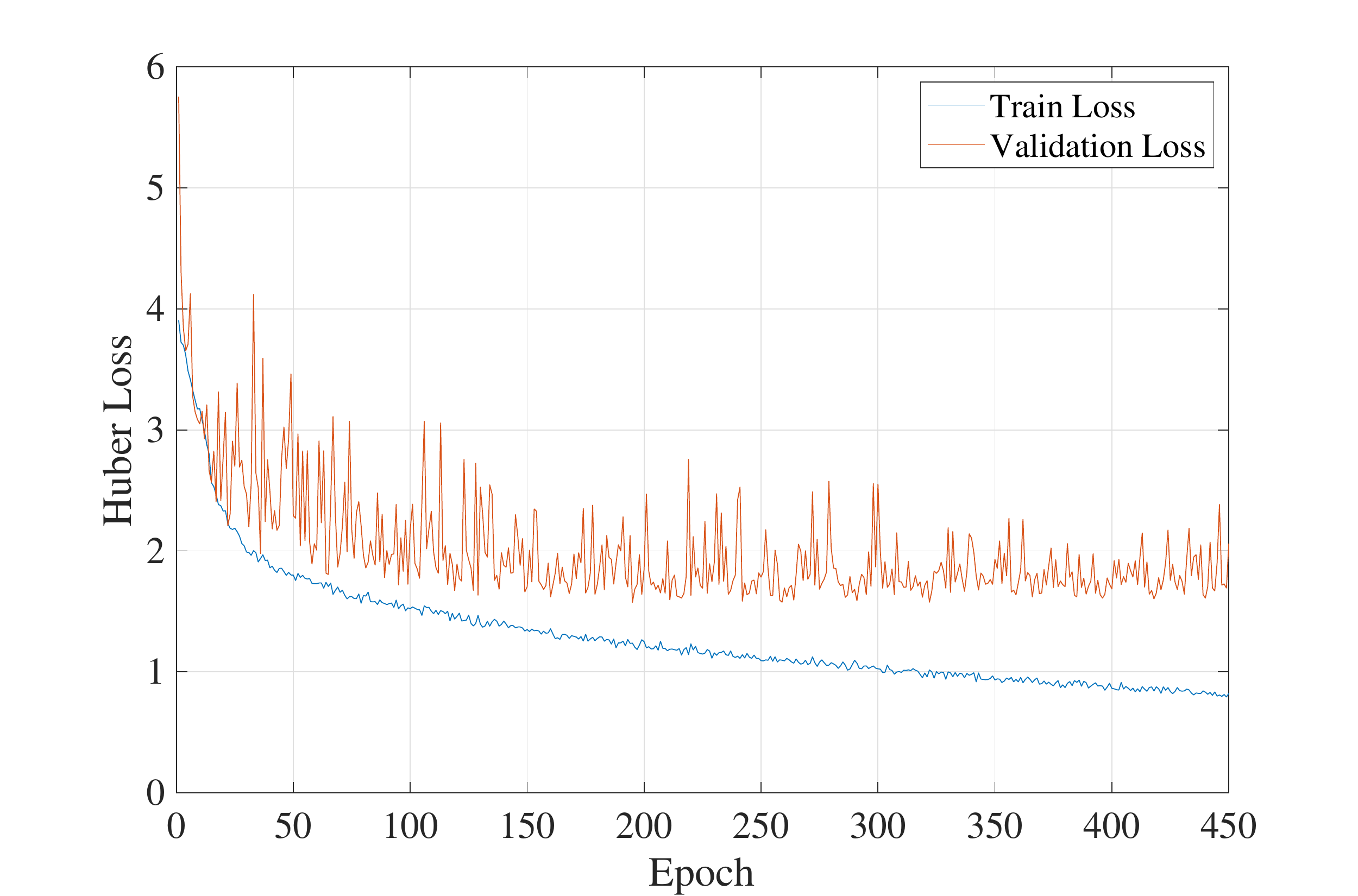}
				\label{fig:epoch_loss_age}
			}
			\subfigure[]{
			\includegraphics[width=0.45\textwidth]{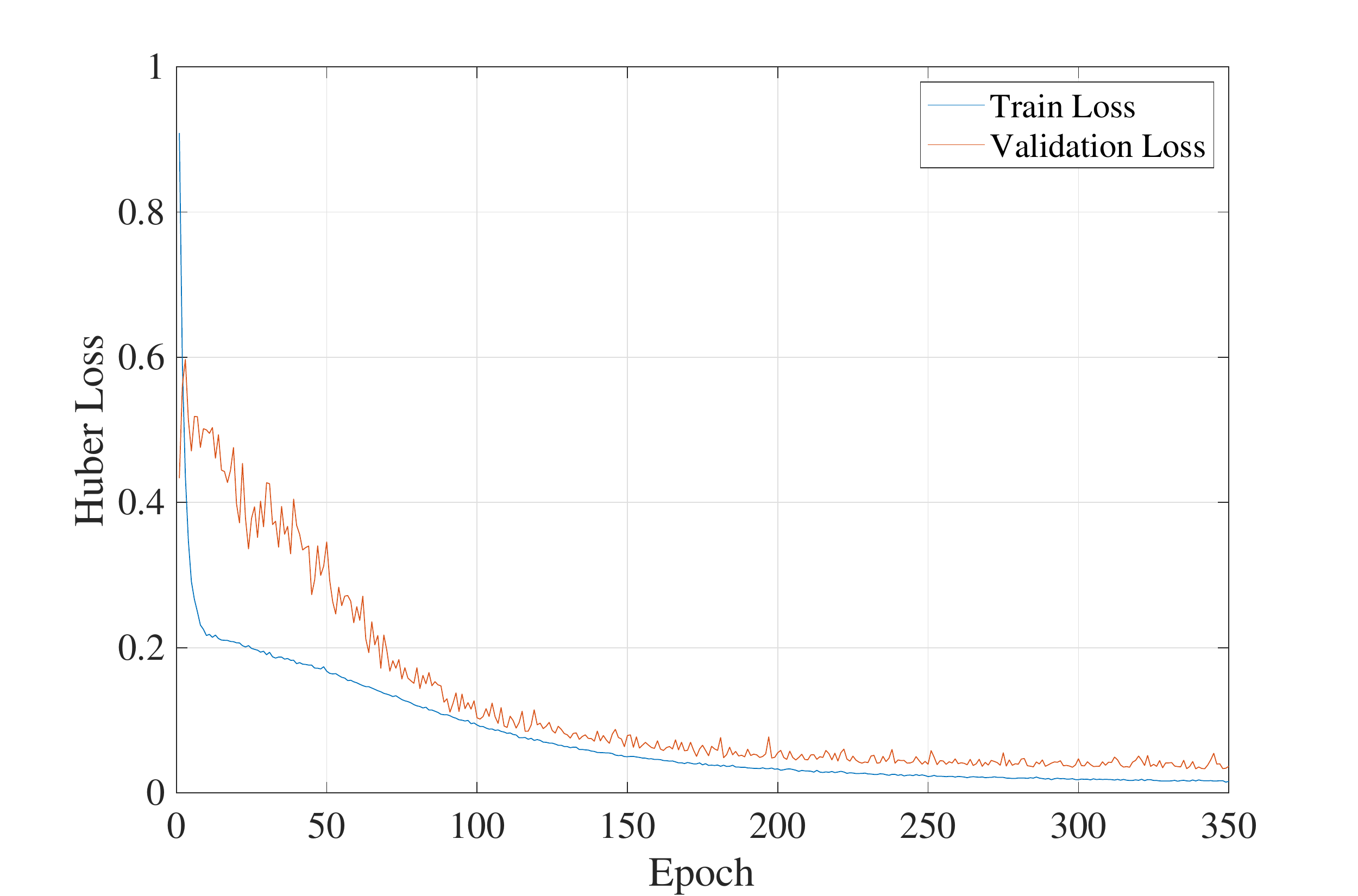}
				\label{fig:epoch_loss_mass}
			}
			\caption{Changes of Huber loss on both training set and validation set during a training (among 10) of age estimation model (a) and mass estimation model (b). The blue and orange curves represent the change of Huber loss on training set and validation set, respectively. As the number of epochs increase, the training loss decreases smoothly, whereas the decline of validation loss is accompanied by oscillations with gradually decreasing amplitude. In this training of model estimating ages, over-fitting appears after approximately 200 epochs and the weights trained with 235 epochs are chosen. Over-fitting appears earlier in the training of mass estimation model, at approximately 150 epochs, and the weights trained with 163 epochs are chosen.}
		\end{figure}
	\subsection{Age Estimation}
	\textbf{
	        Considering the random initialization of the parameters and the use of a dropout layer in training could bring the uncertainty to the model, to improve the generalization capability of our model, we train 10 times on the same training set with different parameter initialization, which is known as model averaging \citep{hansen1990neural}. Each training process contains 450 epochs. Figure \ref{fig:epoch_loss_age} shows the variance of Huber loss as the number of epochs increases in each training process. For each model, we choose the weights allowing the model to have the minimum $M_2$ on the validation set. 
		}
		
\begin{figure}[htbp]
				\centering
				\subfigure[]{
					\includegraphics[width=0.45\textwidth, trim={3cm 0 3cm 0}, clip]{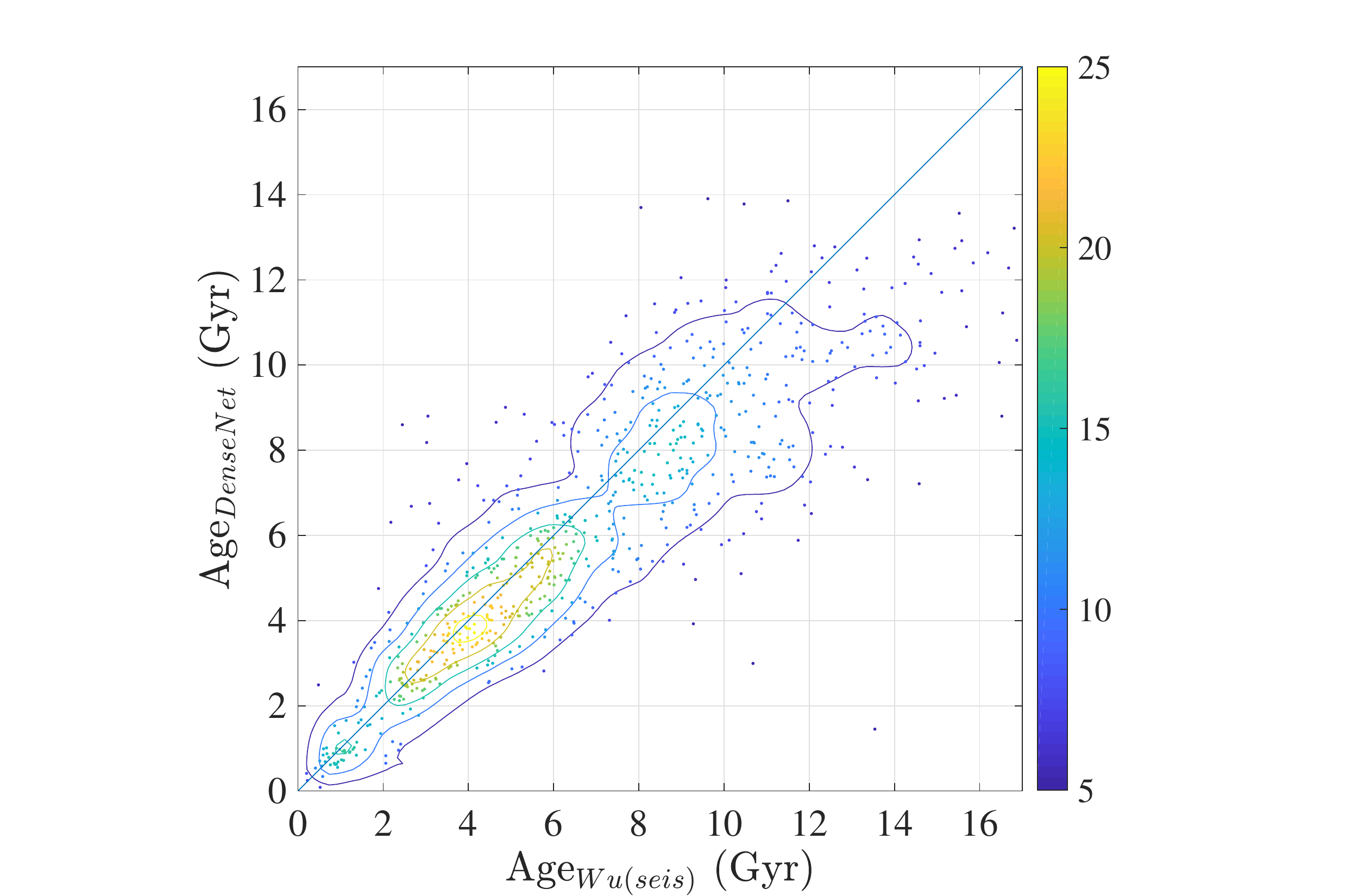}
					\label{fig:pred_true_age}
				}
				\subfigure[]{
					\includegraphics[width=0.45\textwidth, trim={3cm 0 3cm 0}, clip]{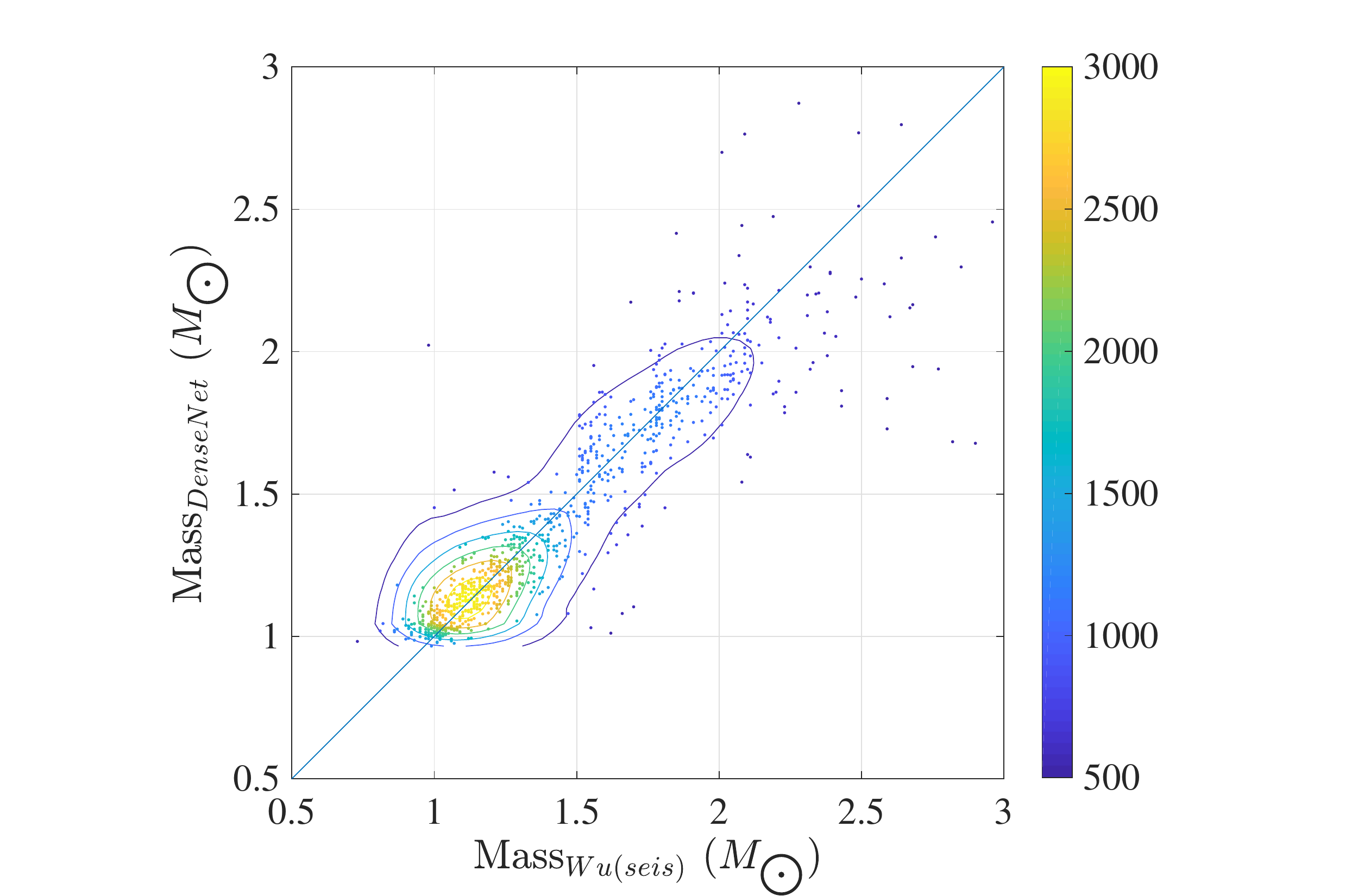}
					\label{fig:pred_true_mass}
				}
				\caption{Left figure demonstrates the prediction given by averaged DenseNet model versus the input values of validation data. Each point is color-coded with density. The closed curves are the density contours. The grey solid line represents the 1:1 line. The right figure is the same as the left, but for the mass estimate.}
			\end{figure}

		\textbf{	
			Then, the data in the validation set are put into these 10 trained networks to get corresponding predictions, after which we compare them with the true values (see Figure \ref{fig:pred_true_age}). The $M_1$ of each model is in the range of 1.51 to 1.63, with a mean of 1.56, and the  $M_2$ ranges from 24.4\% to 25.8\% with a mean of 25.1\%. The standard deviation of the predicted ages from each model is adopted as the uncertainty at a level of 10\%. The $M_1$ and $M_2$ of predictions of the averaged model and the ages in validation set are 1.47 and 23.4\%, respectively.
			}

	\subsection{Mass Estimation}
	\textbf{
	    For mass estimation, we use the same method to obtain the averaged results. We find that the training process converges faster than that in age estimation, and thus this model is trained with 350 epochs in each of 10 training processes as shown in Figure \ref{fig:epoch_loss_mass}. The selection of weights of each model is the same as that in age estimation.  
		}
		
		\textbf{
		As in the age estimation, we compare the mass predictions given by each model with values in validation set (see Figure \ref{fig:pred_true_mass}). The $M_1$ of each model is  in a range of {0.09 to 0.10} and the  $M_2$ is from {6.7\% to 7.1\%} with a mean of {7.0\%}. The uncertainty is {2.5}\%. The $M_1$ and $M_2$ of predictions of the averaged model and the masses in validation set are {0.09 and 6.5\%}, which indicates that our method can estimate the masses of RGB stars accurately. }
		\begin{figure}[htbp]
			\centering
			\subfigure[]{
				\includegraphics[width=0.45\textwidth]{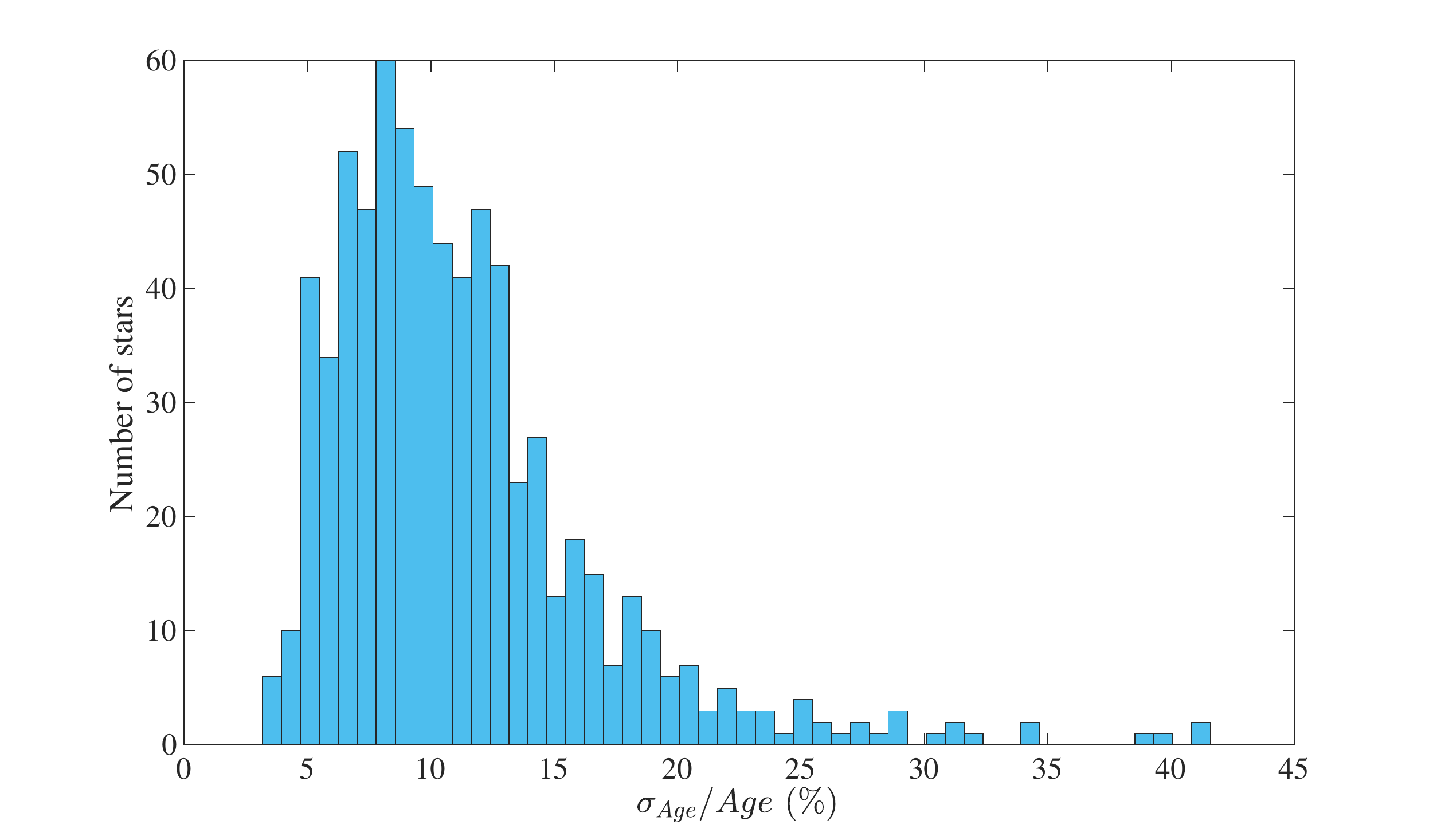}
				\label{fig:uncertainty_age}
			}
			\subfigure[]{
				\includegraphics[width=0.45\textwidth]{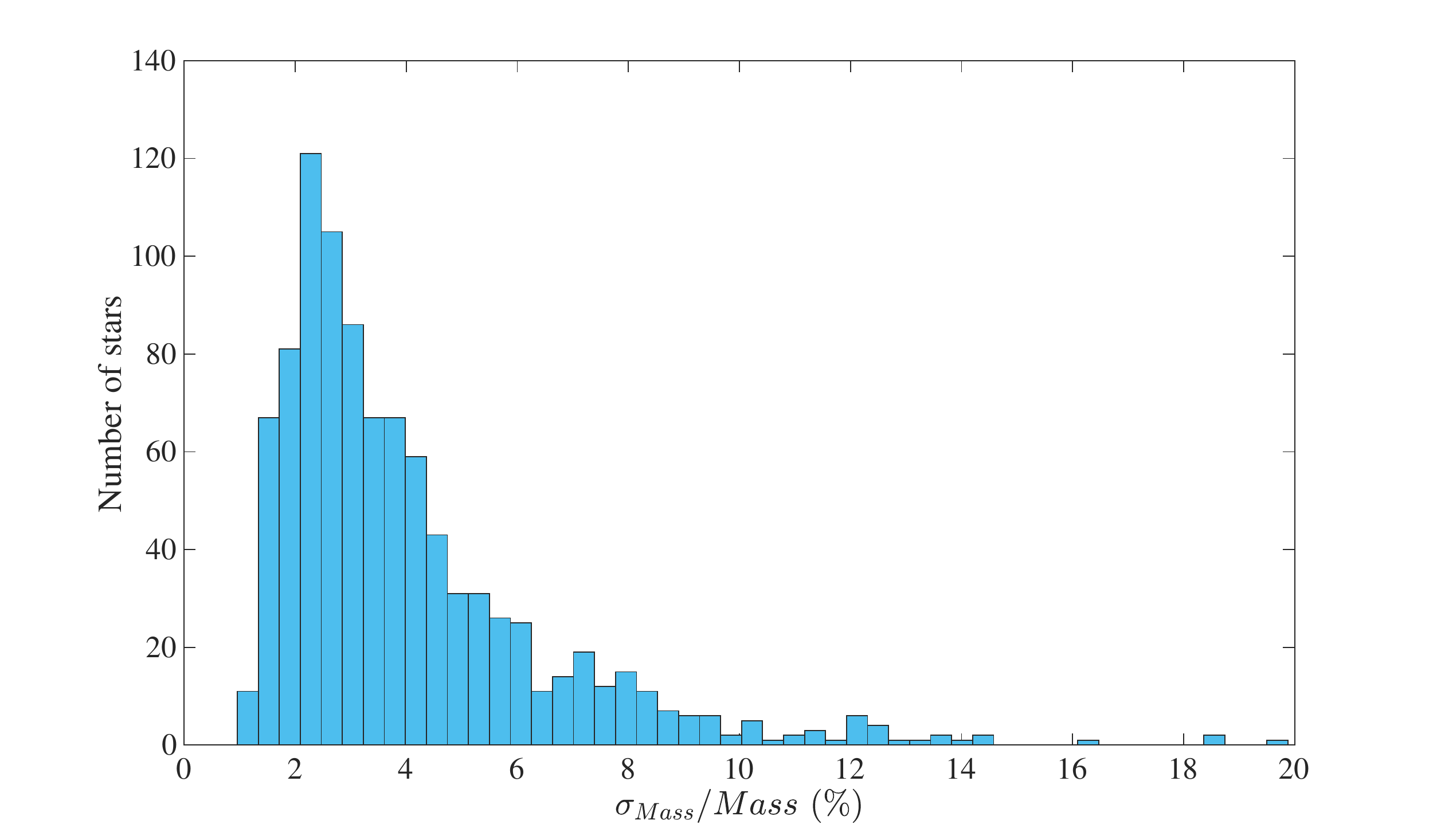}
                \label{fig:uncertainty_mass}
			}
			\caption{Distribution of the fractional uncertainties of  DenseNet model’s prediction of (a) ages and (b) masses.}
		\end{figure}
\section{Discussion}\label{discussion}
    \subsection{Classification of RGB and RC} \label{sec:classificaiton}
    \textbf{
    Core-helium burning RC is a different evolutionary state from hydrogen-shell burning RGB, but stars of these two evolutionary states could appear in the same region in the HRD. Because RC stars have an uncertain amount of mass loss, which may reduce the accuracy of inferred ages, we used a pure RGB dataset to train and validate in the age estimation. However, distinguishing the RC stars from RGB stars is considerably difficult. Here, we introduce a classification model to identify RGB stars from the RC stars.
    }
    
    \textbf{
    A widely used method for distinguishing RGB and RC stars uses asteroseismology.
    The solar-like oscillations have acoustic (p-mode) and gravity (g-mode) characteristics.
    The observed stellar pulsations of evolved stars include information of both p- and g- modes. With the same luminosity and radius, RC stars have a remarkably stronger coupling between g- and p-modes, which leads to the larger period spacing.
    Hence, the large frequency spacing ($\Delta\nu$) and period spacing ($\Delta P$) offer a significantly clean separation of RC stars and RGB stars. \citet{vrard2016period} provide an accurate large frequency spacing and period spacing for thousands of stars and conform to previous measurements including evolutionary states in red giant Kepler public data. We perform the cross-identification of the sample from \citet{vrard2016period} with LAMOST-Kepler stars, obtaining 6\,800 spectra with spectral SNRs higher than 15, including 2\,345 RGB stars and 4\,455 RC stars. As in the process of splitting the datasets, we divide these spectra data into two parts (training and validation sets) in the ratio of 8:2. In the training set, we have 1\,467 RGB stars and 3\,599 RC stars, and the validation set consists of 878 RGB stars and 856 RC stars. To further test the model, we use stars with accurate and clear evolutionary states in the sample of \citet{pinsonneault2018second}. After the cross-identification with LAMOST spectra, we then select the stars that do not appear in \citet{vrard2016period}, including 2\,575 RGB stars and 807 RC stars, as an independent test set.}
    
    \textbf{
    The structure of the model remains unchanged except that in the output layer, the activation function is changed from ReLU to Softmax. Softmax is a generalization of the logistic function to multiple dimensions, which gives the probability for the occurrence of each class j. Here, j is either an RGB (class 0) or an RC (class 1), and thus the output is given by
    \begin{equation}
        p(y=j|\textbf{x}) = \frac{e^{x\cdot \omega_j}}{\sum_{k=1}^2 e^{x\cdot \omega_k}},
    \end{equation}
    where $\bm{x}$ are input values to the output layer and $\bm{\omega}$ are the weights of the output layer. $p$ values closer to 0 or 1 represent higher confidence in classifying an RGB star or an RC star, respectively. The loss function in the training process is changed to cross-entropy loss function given as}
    \begin{equation}
        E(\bm{y}, \bm{\hat{y}}) = -\frac{1}{m}\sum_{i=1}^m[y_i\log \hat{y}_i + (1-y_i)\log (1-y_i)],
    \end{equation}
    \textbf{where $y$ is the ground truth label, $\hat{y}$ is the predicted probability, and $m$ is the number of spectra in the training set. The optimizer is still Adam with the learning rate being 0.0005. The method of initializing weights is the He uniform variance scaling initializer.}
    
\begin{figure*}[!h]
		\centering
		\includegraphics[width=0.7\linewidth]{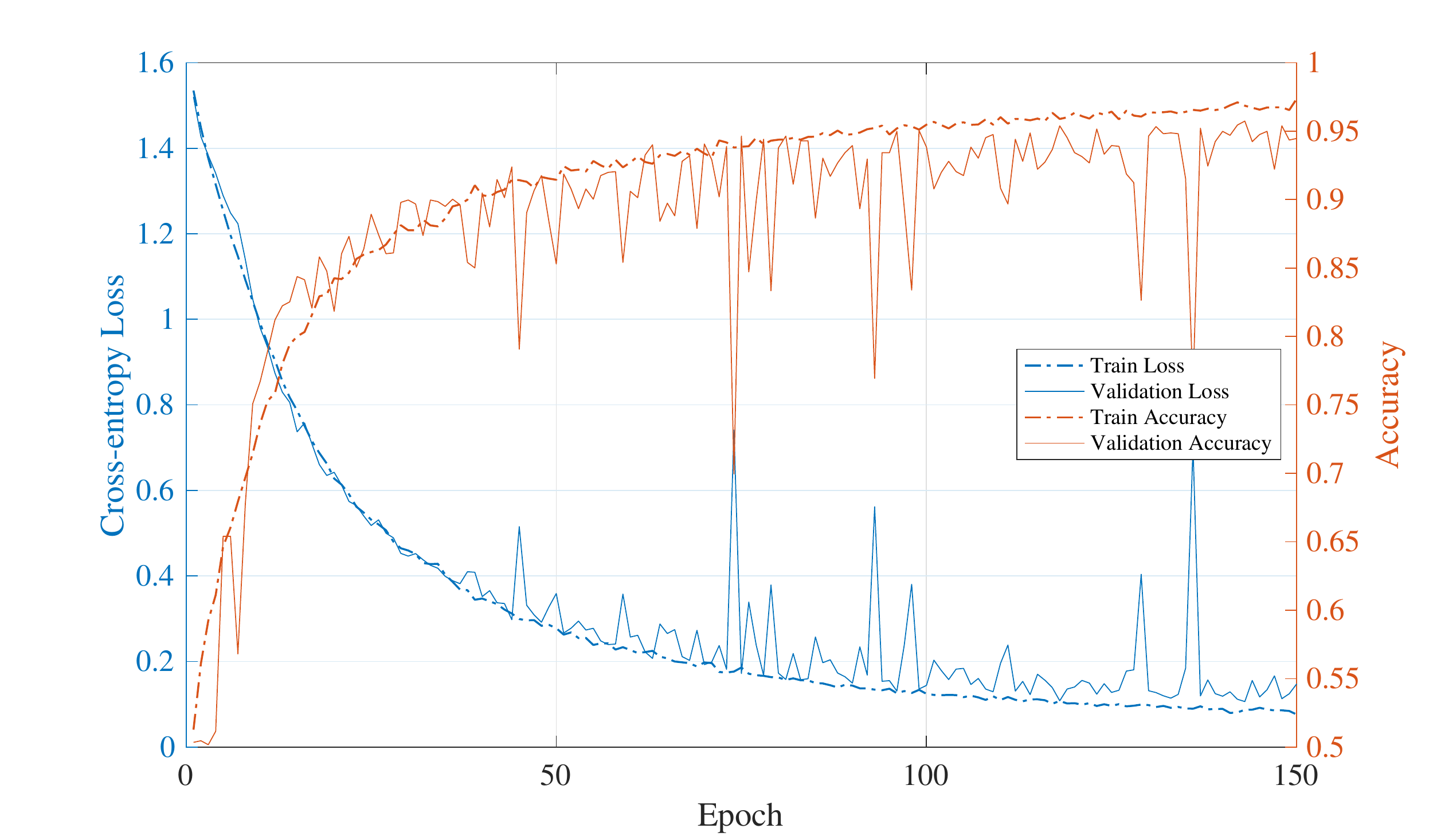}
		\caption{Variance of cross-entropy loss (blue) and accuracy (orange, means the ratio of all correct predictions to all predictions) on the training set (solid line) and validation set (dot dash line) in one of ten training processes.}
		\label{fig:loss_acc}
	\end{figure*}

    \textbf{The averaging method introduced above is adopted. We train 150 epochs each time and select the one with minimum loss. Figure \ref{fig:loss_acc} shows one of ten classification model training processes.}
	
    \textbf{We use the averaged model to identify RGB stars from the validation set and determine that the contamination rate is 0.4\% and the completeness is 98.0\%. In the test set, we find the contamination rate ranges from 1.2\% to 2.6\% at a level of 1.8\% and the completeness is $\sim$ 70\%. Using the averaged model, the contamination and completeness are 0.9\% and 66.2\%. The contamination rate is lower than that of methods in the literature based on stellar atmospheric parameters alone \citep{bovy2014apogee,huang2015metallicity}. A further improvement should expand the training set and adjust the network structure appropriately to improve the completeness, which can select more RGB stars and maintain the purity at the same time.}
    \subsection{Test  on the  Open  Clusters }
  Stars in the same cluster are universally believed to form at almost the same time and location with the same age.
		To further verify the accuracy, robustness, and generalizability of our model, we tested it by using the RGB stars' spectra of three different open clusters, M 67, Berkeley 32, and NGC 2420.    \textbf{The member-star identification for these open clusters is presented in \citet{xiang2017estimating}. From their work, 88, 2858, and 72 membership candidates for NGC 2420, M67, and Berkeley 32, respectively, were identified. 
We used the method given by \citet{huang2019milky} to select RGB stars and RC stars in these three clusters,
and the selection rules are as follows:}
		\begin{itemize}
			\item The effective temperature of the star is below $10000$ K;
			\item The surface gravity of the star satisfies $\log g \leq 3.5$.
		\end{itemize}
		\textbf{
		We then used the averaged classification model introduced in Section \ref{sec:classificaiton} to identify RGB stars, after which we further processed these data in a similar manner to that used in previous experiments and obtained 36 RGB samples from M67, 15 RGB samples from Berkeley 32, and 22 RGB samples from NGC2420. We then applied our averaged model for age estimation to these samples, and the results are shown in Table \ref{tb:validation result}. It is observed that the ages given by our model are  consistent well with the literature values. }
		\begin{table}[!htbp]
			\centering
			\begin{tabular}{ccccc}
			\toprule
				 Open cluster & Age$^*$ (Gyr) & $\sigma_{Age^*}$ (Gyr) &Age$_{Den}$ (Gyr) & $\sigma_{Age_{Den}}$ (Gyr)\\
			\midrule
				M67 & 4.0 & 3.5--4.8 & 4.1 & 3.9--4.4\\
				Berkeley 32 & 6.0 & 5.0--7.2 & 6.0 & 5.3--6.7\\
				NGC2420 & 3.3 & 2.8--3.8 & 3.2 & 2.8--3.6\\
			\bottomrule
			\end{tabular}
			\caption{Age$_{Den}$ means the ages of open clusters  from DenseNet. The literature ages (Age$^*$) and their uncertainty ($\sigma_{Age^*}$) of M67, Berkeley 32, and NGC 2420 are from \citet{demarque1992solar}, \citet{kaluzny1991photometric}, and \citet{mcclure1974old}, respectively. }
			\label{tb:validation result}
		\end{table}
    \subsection{Comparison with Other Methods}	
		In this section, we demonstrate the effectiveness of our model on the evaluation of ages and masses of RGB stars by comparing it with three other widely used  machine learning methods: XGBoost (XGB), random forest (RF), and ANN. The comparison results are presented in Table \ref{tabel:comparison}.
		The comparison shows that XGB and RF have a similar performance with the errors of $\sim 36\%$ and $\sim 12\%$ on age estimation and mass estimation, respectively. The other deep learning method, ANN, performs worst among these four methods, mainly because the dimension of the input spectral data is too high. Our model performs better than other machine learning methods in estimating the ages and masses of stars.
		\begin{table}[!htbp]
			\centering
			\begin{tabular}{ccccc}
			\toprule
			\multirow{2}{*}{Model} & \multicolumn{2}{c}{Age Results}& \multicolumn{2}{c}{Mass Results}\\
			\cline{2-3} \cline{4-5} 
			  & $M_1$ & $M_2$ & $M_1$ & $M_2$\\
			\midrule
				XGB& 1.65 & 36.0$\%$ & 0.20 & 13.4$\%$\\
				RF & 1.74 & 36.5$\%$ & 0.18  & 11.7$\%$\\
				ANN & 5.08 &  67.3$\%$ & 0.51 & 33.2$\%$\\
				DenseNet-BC & 1.47 & 23.4$\%$ &  {0.09} &  {6.5}$\%$\\
			\bottomrule
			\end{tabular}
			\caption{Comparison of different machine learning models  in estimating the ages and masses of stars. XGB: XGBoost, RF: Random forest, ANN: Artificial neural network.}
			\label{tabel:comparison}
		\end{table}
	\subsection{Comparison with Other Ages and Masses}
\begin{figure}[htbp]
		\centering
		\subfigure[]{
			\includegraphics[width=0.45\textwidth]{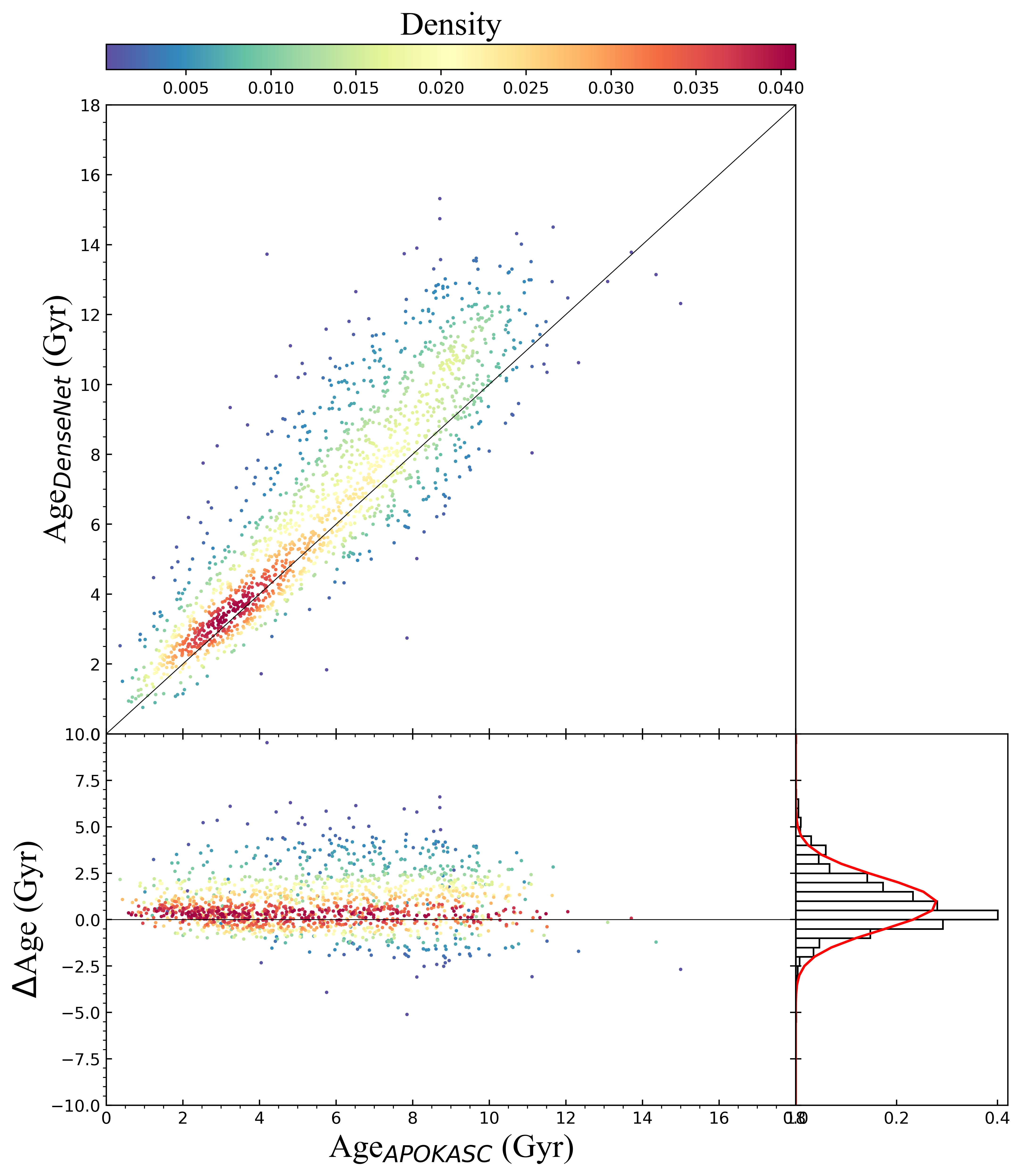}
		}
		\subfigure[]{
			\includegraphics[width=0.45\textwidth]{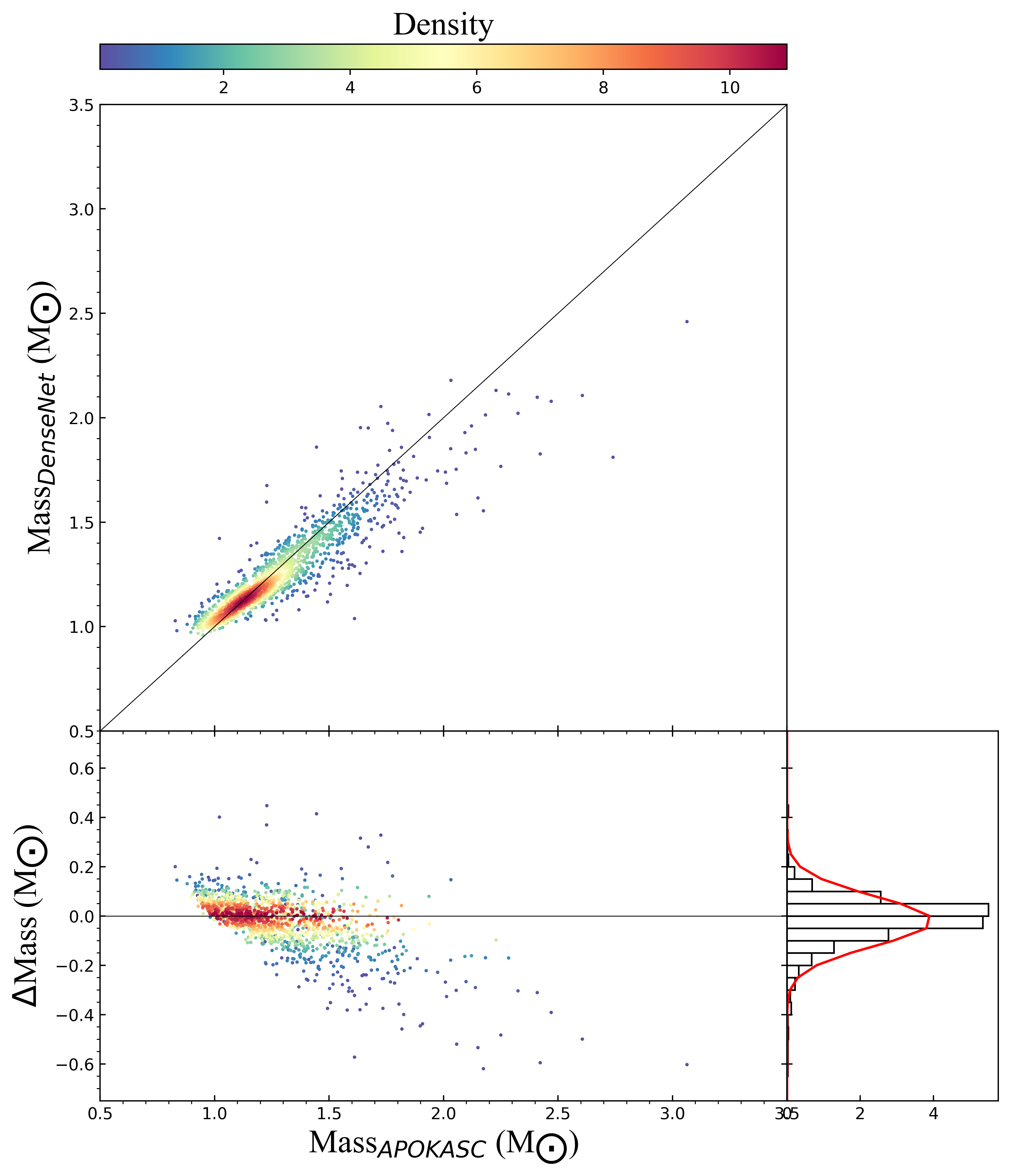}
		}
		\caption{Comparison between the (a) age and (b) mass given by averaged DenseNet model and  APOKASC-2 catalog \citep{pinsonneault2018second} with the residual distribution.}
		\label{fig:apo_our}
	\end{figure}

	\textbf{
	We first compare the ages and masses derived from DenseNet with APOKASC-2, as shown in Figure \ref{fig:apo_our}. Owing to the addition of subgiant stars in our training process, compared with the difference between the estimation of \citet{wu2018mass} and \citet{pinsonneault2018second}, 
	we reduce the mean difference of the age estimation to 20\% and maintain the consistency in mass estimation. }
	    
\begin{figure}[!h]
                 \centering
                \includegraphics[width=0.5\textwidth]{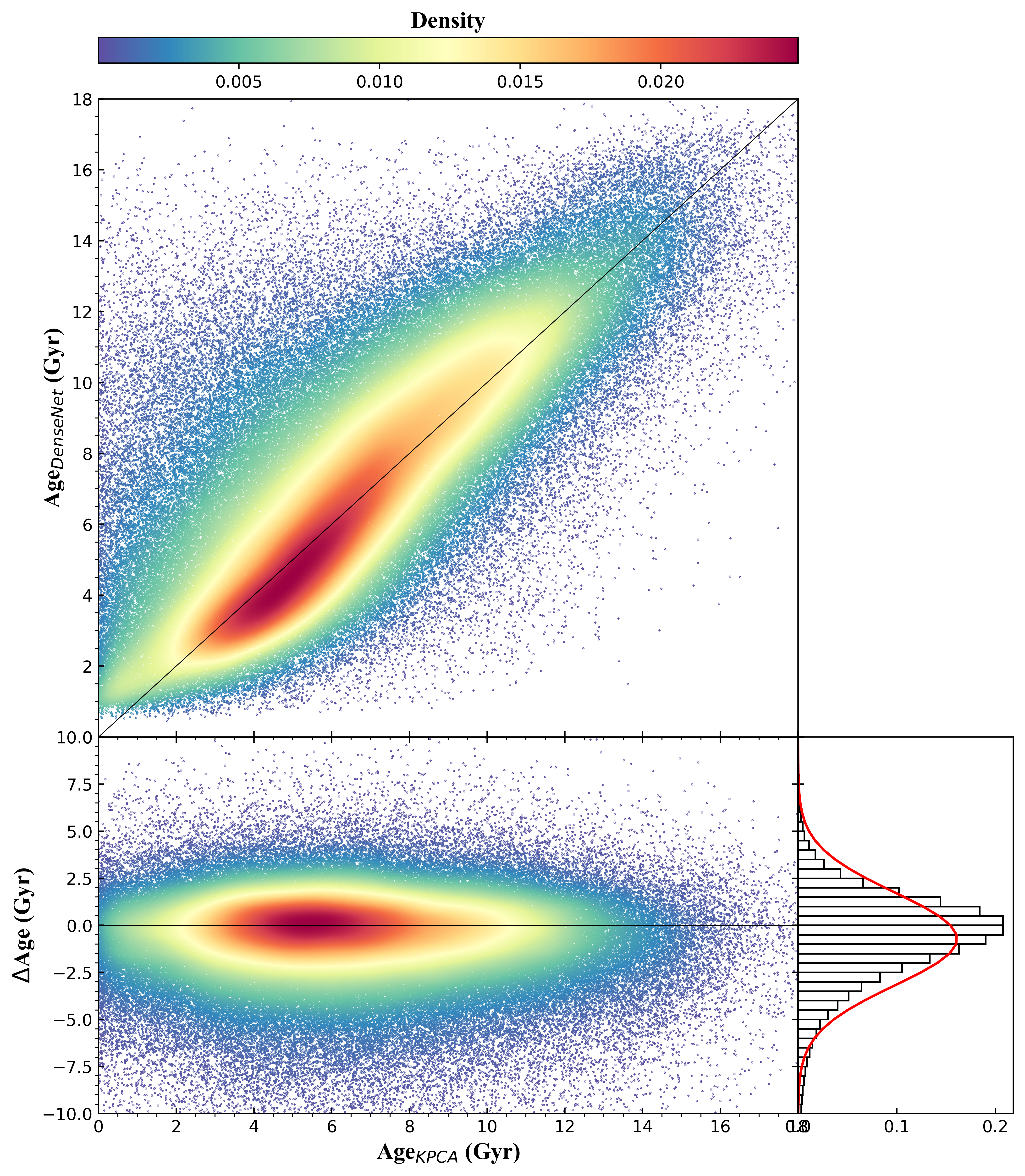}
                 \caption{Upper figure shows the comparison between the  ages given by the averaged DenseNet model and the catalog offered by \citet{wu2019ages} using KPCA. The lower ones indicate the distribution and variance of residuals.}
                  \label{fig:age_kpca_our}
        \end{figure}

         \textbf{
         Moreover, \citet{wu2019ages} provide a catalog of  masses and ages for 480\,000 stars given by KPCA from LAMOST DR4. We cross-matched their catalog with our 512\,272 RGB stars (SNR $> 15$, $\log \mathrm{g} < 3.5,  T_{\mathrm{eff}} < 10000 \mathrm{K}$) from LAMOST DR7, obtaining 210\,000 common RGBs. Figure \ref{fig:age_kpca_our} plots the comparison between the ages given by \citet{wu2019ages} and our method. The median difference between these two  works is approximately 0.43 Gyr.}

\section{Conclusion}
	 In this paper, we proposed a new data-driven model based on the   deep learning method, DenseNet, to determine the ages of RGB stars. 
	 In the experiment, we first apply wavelet analysis to denoise the spectra and then we apply DenseNet to the reshaped data to determine the ages and masses of RGB stars. The results show that our method can determine the ages with an accuracy of 24.3$\%$ and mass with an accuracy of {6.5$\%$}. \textbf{Then, we use the structure of the regression model to build a classifier to select RGB stars. The contamination rate on the test set is only $\sim$ 1.8\%. In both the regression model and classification model, we use “model averaging” method to evaluate the uncertainty and give more robust results. We have applied the method to all the RGB stars selected from LAMOST DR7, and the result is provided as a supplemental catalog (Doi: 10.5281/zenodo.4640118), which can be accessed through   \url{https://github.com/xhRhapsody/LAMOST_DR7_RGB_DenseNet_MassAge} and \url{ https://zenodo.org/record/4640118}.} The catalog provides a good sample  for future studies. 
	 
 \section{Acknowledgments}
This work is supported by National Natural Science Foundation of China under grant numbers  11873037, 11603012, and partially supported by the Young Scholars Program of Shandong University, Weihai (2016WHWLJH09), Natural Science Foundation of Shandong Province, China
(ZR2015AQ011), and China postdoctoral Science Foundation (2015M571124).  

\section{Software}
\software{
\begin{enumerate}
    \item DenseNet$\ $\citep{huang2017densely, huang2019convolutional}:$\ $ \url{https://github.com/liuzhuang13/DenseNet };
    \item XGBoost:$\ $\url{https://github.com/dmlc/xgboost };
    \item Scikit-learn$\ $\citep{pedregosa2011scikit}:$\ $\url{https://scikit-learn.org/stable/modules/generated/sklearn.ensemble.RandomForestRegressor.html}.
\end{enumerate}
}
\newpage

\bibliographystyle{aasjournal}
\bibliography{ref}

\end{document}